\documentclass{aa}  

\usepackage{graphicx}
\usepackage[dvipsnames]{xcolor}
\usepackage[varg]{txfonts}
\usepackage[colorlinks=true,linkcolor=Bittersweet,citecolor=Bittersweet,urlcolor=ForestGreen]{hyperref} 
\usepackage{enumerate}
\usepackage{amsmath}
\usepackage{natbib}
\bibpunct{(}{)}{;}{a}{}{,}
\usepackage{comment}
\usepackage[para]{threeparttable}
\usepackage{multirow}
\usepackage{caption}
\usepackage{floatrow} 
\usepackage{subcaption}
\usepackage{float}
\usepackage{placeins}
\usepackage{afterpage}
\usepackage{array,booktabs}
\usepackage{siunitx}
\begin{document}

\title{High-resolution, 3D radiative transfer modelling}
\subtitle{V. A detailed model of the M~51 interacting pair.}

\author{
Angelos Nersesian\inst{1} \and
S\'{e}bastien Viaene\inst{2} \and
Ilse De Looze\inst{2,3} \and
Maarten Baes\inst{2} \and
Emmanuel M. Xilouris\inst{1} \and
Matthew W. L. Smith\inst{4} \and
Simone Bianchi\inst{5} \and
Viviana Casasola\inst{6} \and
Letizia P. Cassar\`{a}\inst{7}\and
Christopher J. R. Clark\inst{8} \and
Wouter Dobbels\inst{2} \and
Jacopo Fritz\inst{9} \and
Fr\'{e}d\'{e}ric Galliano\inst{10} \and
Suzanne C. Madden\inst{10} \and
Aleksandr V. Mosenkov\inst{11,12} \and
Ana Tr\v{c}ka\inst{2}
}

\institute{National Observatory of Athens, Institute for Astronomy, Astrophysics, Space Applications and Remote Sensing, Ioannou Metaxa and Vasileos Pavlou GR-15236, Athens, Greece \\ \email{\textcolor{blue}{a.nersesian@noa.gr}} \and  
Sterrenkundig Observatorium Universiteit Gent, Krijgslaan 281 S9, B-9000 Gent, Belgium \and
Department of Physics and Astronomy, University College London, Gower Street, London WC1E 6BT, UK \and
School of Physics and Astronomy, Cardiff University, The Parade, Cardiff CF24 3AA, UK \and
INAF - Osservatorio Astrofisico di Arcetri, Largo E. Fermi 5,I-50125, Florence, Italy \and
INAF - Istituto di Radioastronomia, Via P. Gobetti 101, 4019, Bologna, Italy \and
INAF - Istituto di Astrofisica Spaziale e Fisica Cosmica, Via Alfonso Corti 12, 20133, Milan, Italy \and
Space Telescope Science Institute, 3700 San Martin Drive, Baltimore, Maryland, 21218, USA \and
Instituto de Radioastronom\'{i}a y Astrof\'{i}sica, UNAM, Campus Morelia, AP 3-72, 58089 Michoac\'{a}n, Mexico \and
Laboratoire AIM, CEA/DSM - CNRS - Universit\'{e} Paris Diderot, IRFU/Service d’Astrophysique, CEA Saclay, 91191, Gif-sur- Yvette, France \and
Central Astronomical Observatory of RAS, Pulkovskoye Chaussee 65/1, 196140, St. Petersburg, Russia \and 
St. Petersburg State University, Universitetskij Pr. 28, 198504, St. Petersburg, Stary Peterhof, Russia
}

\date{Received 16 July 2020 / Accepted  10 September 2020}


\abstract{
    Investigating the dust heating mechanisms in galaxies provides a deeper understanding of how the internal energy balance drives their evolution. Over the last decade, radiative transfer simulations based on the Monte Carlo method have underlined the role of the various stellar populations heating the diffuse dust. Beyond the expected heating through ongoing star formation, both older stellar population ($\ge 8~$Gyr) and even AGN can contribute energy to the infrared emission of diffuse dust.}
    {
    In this particular study, we examine how the radiation of an external heating source, like the less massive galaxy NGC~5195, in the M~51 interacting system, could affect the heating of the diffuse dust of its parent galaxy, NGC~5194, and vice versa.}
    {
    We use \textsc{SKIRT}, a state-of-the-art 3D Monte Carlo radiative transfer code, to construct the 3D model of the radiation field of M~51, following the methodology defined in the DustPedia framework. In the interest of modelling, the assumed centre-to-centre distance separation between the two galaxies is $\sim10~$kpc.}
    {
    Our model is able to reproduce the global spectral energy distribution of the system, and it matches the resolved optical and infrared images fairly well. In total, 40.7\% of the intrinsic stellar radiation of the combined system is absorbed by dust. Furthermore, we quantify the contribution of the various dust heating sources in the system, and find that the young stellar population of NGC~5194 is the predominant dust-heating agent, with a global heating fraction of 71.2\%. Another 23\% is provided by the older stellar population of the same galaxy, while the remaining 5.8\% has its origin in NGC~5195. Locally, we find that the regions of NGC~5194 closer to NGC~5195 are significantly affected by the radiation field of the latter, with the absorbed energy fraction rising up to 38\%. The contribution of NGC~5195 remains under the percentage level in the outskirts of the disc of NGC~5194. This is the first time that the heating of the diffuse dust by a companion galaxy is quantified in a nearby interacting system.}
{}


\maketitle
%

\section{Introduction} \label{sec:intro}

Galaxy interactions are extremely powerful mechanisms that drive the formation and evolution of galaxies in the Universe \citep{Toomre_1972ApJ...178..623T, Toomre_1977egsp.conf..401T, Somerville_1999MNRAS.310.1087S, Kauffmann_1999MNRAS.303..188K, Kauffmann_1999MNRAS.307..529K, Springel_2005Natur.435..629S}. It is well known that interacting galaxies have higher dust and gas contents than unperturbed galaxies \citep[e.g.][]{Combes_1994A&A...281..725C, Casasola_2004A&A...422..941C, Violino_2018MNRAS.476.2591V, Lisenfeld_2019A&A...627A.107L, Moreno_2019MNRAS.485.1320M}. Due to the strong tidal forces generated by such interactions, the redistribution and accumulation of baryonic matter (gas and dust inflows) in the interstellar medium (ISM) becomes possible, which triggers and/or enhances the formation of new stars \citep{Kennicutt_1998ARA&A..36..189K, Kauffmann_2000MNRAS.311..576K, Woods_2007AJ....134..527W}. The interaction between galaxies may be even responsible, to some extent, for the variety of morphologies and morphological distortions that we observe. For example, numerical simulations strongly indicate that most elliptical galaxies are the end product of galaxy merging \citep[e.g.][and references therein]{Toomre_1972ApJ...178..623T, Toomre_1977egsp.conf..401T, Cox_2006ApJ...650..791C, Cox_2006MNRAS.373.1013C, Naab_2006MNRAS.369..625N}. 

Depending on the mass ratio of the involved galaxies, mergers are categorised into major and minor. The properties of mergers have been well studied, both observationally \citep[e.g.][]{Woods_2007AJ....134..527W, Ellison_2008AJ....135.1877E, Ellison_2011MNRAS.418.2043E, Ellison_2013MNRAS.435.3627E, Weston_2017MNRAS.464.3882W}, and with numeric simulations \citep[e.g.][]{Toomre_1972ApJ...178..623T, Barnes_1992ARA&A..30..705B, Barnes_1996ApJ...471..115B, Mihos_1996ApJ...464..641M, Di_Matteo_2007A&A...468...61D, Cox_2008MNRAS.384..386C}. Major mergers (i.e. galaxies of approximately equal-mass) are responsible for the brightest objects in the local Universe, the ultraluminous infrared galaxies (ULIRGs), with bolometric luminosities greater than $10^{12}~\text{L}_\odot$ \citep[e.g.][]{Sanders_1996ARA&A..34..749S, Clements_1996MNRAS.279..477C, Dasyra_2006ApJ...638..745D}. ULIRGs are characterised by enormous levels of ongoing star-formation activity, but they are also heavily obscured by dust, emitting the bulk of their energy in the far-infrared (FIR) wavelengths. Moreover, the interactions between galaxies of unequal mass (i.e. minor mergers) are also an integral component of galaxy evolution, and could potentially lead towards an enhancement in star formation \citep{Woods_2007AJ....134..527W}. \citet{Kaviraj_2014MNRAS.440.2944K} showcased the importance of minor mergers in the evolution of massive galaxies in the local Universe, with about half of the star-formation activity driven by their merging process.

One of the most famous mergers in the local Universe is that of the grand-design spiral arm galaxy NGC~5194 (M~51a or the \textit{Whirlpool} galaxy), and its companion, lenticular galaxy NGC~5195 (M~51b). The morphological type of M~51a is Sbc, and that of M~51b is SB0 \citep{de_Vaucouleurs_1995yCat.7155....0D}. The centre of M~51a harbours a weak active galactic nucleus (AGN) classified as a Seyfert 2 \citep{Terashima_2001ApJ...560..139T, Bradley_2004ApJ...603..463B}, while M~51b is classified as a LINER \citep{Satyapal_2004A&A...414..825S, Moustakas_2010ApJS..190..233M}. Usually, the term `M~51' refers to NGC~5194 only, but in order to avoid confusion we will use the term `M~51' to indicate the NGC~5194~$+$~NGC~5195 interacting system (see Fig.~\ref{fig:comp_maps_m51} for an optical image of M~51). An open question that still remains today is whether or not the two galaxies had a single or multiple encounters. Evidence exists in support of either model, based on previous kinematic studies \citep{Durrell_2003ApJ...582..170D, Salo_2000MNRAS.319..377S}, N-body simulations \citep{Salo_2000MNRAS.319..377S, Theis_2003Ap&SS.284..495T}, and hydrodynamic simulations \citep{Dobbs_2010MNRAS.403..625D}. According to the single-encounter model, M~51b crossed M~51a about 300-500~Myr ago, and is currently located at a distance of 50~kpc beyond M~51a \citep{Toomre_1972ApJ...178..623T}. On the other hand, the dynamical models presented in \citet{Salo_2000MNRAS.319..377S} are in favour of a multiple-encounter scenario, suggesting that after the first encounter 400-500~Myr ago, M~51b passed near M~51a for the second time about 50–100 Myr ago. This model places M~51b at a distance of 20~kpc beyond M~51a. Either way, at least one encounter occurred between the two galaxies, causing a burst of star formation 250–550~Myr ago in both of them \citep{Tikhonov_2009AstL...35..599T}, and creating a stellar bridge that connects the galaxies together. In this paper, we adopt the multiple encounter scenario as the most up-to-date representation of M~51 in the literature which successfully explains detailed kinematics based on H\textsc{i} radio observations \citep{Salo_2000MNRAS.319..377S}, and because it is supported by N-body \citep{Salo_2000MNRAS.319..377S, Theis_2003Ap&SS.284..495T} and hydrodynamic simulations \citep{Dobbs_2010MNRAS.403..625D}.

The star formation in M~51a has continued until the present day and is ubiquitous across its disc \citep{Tikhonov_2009AstL...35..599T, Lee_2011ApJ...735...75L}, while M~51b shows a deficiency in ionised gas \citep{Thronson_1991MNRAS.252..550T} and gravitationally stable molecular gas \citep{Kohno_2002PASJ...54..541K}, indicating the absence of ongoing star formation. Yet recent radio observations claim otherwise, i.e. the process of star formation has returned at a low rate in the central region of M~51b \citep{Alatalo_2016ApJ...830..137A}. \citet{Tikhonov_2009AstL...35..599T} determined the stellar composition of M~51b with optical colours, and found M~51b is surrounded by a significant number of asymptotic giant branch (AGB) stars, a result of the starburst 250–550~Myr ago. Furthermore, \citet{Mentuch_Cooper_2012ApJ...755..165M} investigated the dust content of M~51, on spatial scales of $\sim1$~kpc. The authors reported that a strong interstellar radiation field (ISRF) is heating the dust in M~51b up to temperatures of $\sim30~K$, and proposed three heating mechanisms, including heating from the evolved stellar population, and from an AGN. Upon further investigation, both \citet{Eufrasio_2017ApJ...851...10E} and \citet{Wei_2020arXiv200706231W}, do present evidence of an older stellar population ($>5~$Gyr) heating the dust in M~51b, by investigating the star-formation history of M~51 on spatially resolved scales. Another way to probe the heating mechanisms is by constructing a 3D representation of the radiation field of M~51 and quantify the various dust heating sources in the system. 

Inverse radiative transfer simulations of face-on galaxies have now reached the point where they can be computationally efficient, and where they could provide all the necessary tools for a quantitative study of the mechanisms that power the infrared emission by dust, such as the different stellar populations \citep{Viaene_2017A&A...599A..64V, Williams_2019MNRAS.487.2753W, Thirlwall_2020MNRAS.495..835T, Verstocken_2020A&A...637A..24V, Nersesian_2020A&A...637A..25N} or other heating sources, e.g. AGN \citep{Viaene_2020A&A...638A.150V}. In a first attempt to quantify the dust heating fraction related to the star-formation rate (SFR) in face-on spiral galaxies, \citet{De_Looze_2014A&A...571A..69D} created a 3D radiative transfer model of M~51a (Paper I in this series
of radiative transfer studies). They reveal that the contribution of the older stellar population ($\sim10~$Gyr) to the heating of the diffuse dust in the ISM is quite significant, donating up to 35\% of its total energy budget to that cause. However, the authors of that study have not taken into consideration the possible impact of M~51b to the spectral energy distribution of M~51a, which should be non-negligible due to their close relative distance ($\sim20$~kpc).

In this study, the fifth in the series, we will revisit the radiative transfer model of M~51a \citep{De_Looze_2014A&A...571A..69D}, and update the methodology of constructing such a model according to the pipeline established in \citet{Verstocken_2020A&A...637A..24V}. Within the DustPedia framework \citep{Davies_2017PASP..129d4102D}, the method was optimised and applied to the early-spiral galaxy M~81 \citep[Paper II][]{Verstocken_2020A&A...637A..24V}, to a sample of face-on barred galaxies \citep[Paper III][]{Nersesian_2020A&A...637A..25N}, and to NGC~1068 which harbours an AGN \citep[Paper IV][]{Viaene_2020A&A...638A.150V}. Two additional studies employed our modelling strategy to build the models of the large Local Group galaxies M~31 \citep{Viaene_2017A&A...599A..64V} and M~33 \citep{Williams_2019MNRAS.487.2753W}. 

\begin{figure*}[t!]
\centering
\includegraphics[width=\textwidth]{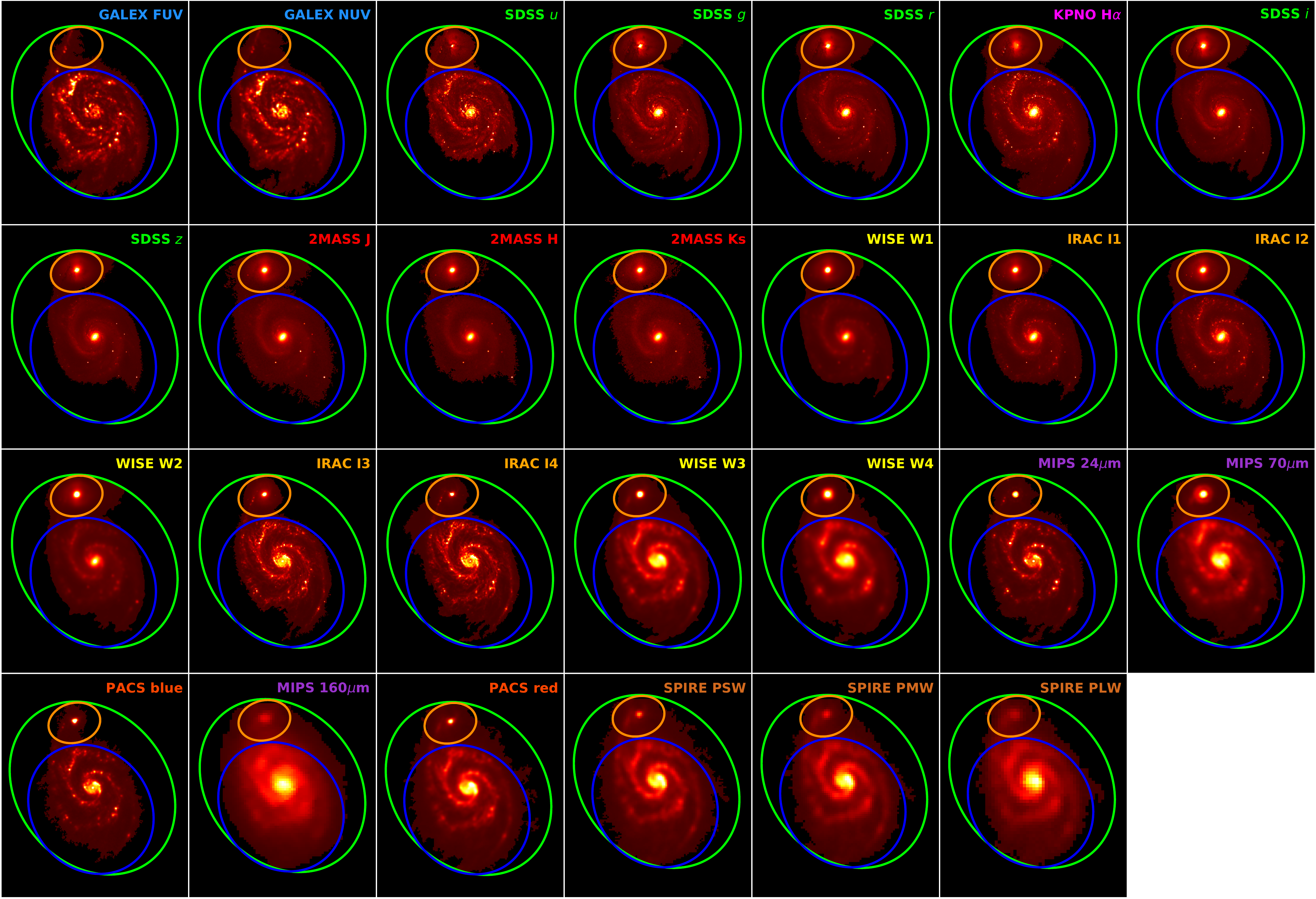}
\caption{Photometry thumbnail image grid for M~51. The green ellipse indicates the master aperture used for the determination of the integrated flux of M~51 and is centred on M~51a. The blue and orange ellipses mark the regions used for the determination of the integrated fluxes of M~51a and M~51b, respectively.}
\label{fig:photometry_M51}
\end{figure*}

Due to its close proximity ($\sim 8~$Mpc) and face-on views (inclination $i\sim33~\deg$), M~51 is a system of particular interest for investigating the dust heating mechanisms on spatially resolved scales across the infrared spectrum. The scope of the paper is to assess, possibly for the first time, the effect of the complex radiation field of an interacting pair of galaxies on the dust heating. We will try to quantify the fraction of radiation by the different heating sources (e.g. old and young stellar populations, and M~51b) on spatial scales up to $\sim150$~pc, by treating M~51 as a uniform system. With this study, we want to provide a new way to quantify the typical exchange of radiation between the members of pre-mergers.

This paper is organised as follows. In Sect.~\ref{sec:dataset} we describe the dataset used in this study, and the steps we followed to correct the images for foreground stars and background noise. We also present the photometry, and the results of a global SED fitting for M~51 with \textsc{CIGALE}. Section~\ref{sec:3dmodel} briefly outlines our modelling approach while in Sect.~\ref{sec:results} we present and validate our results. In Sect.~\ref{sec:dust_heating} we show how the contribution of the different stellar populations of M~51 shape its SED, and quantify their contribution to the dust heating. We investigate what the fraction of radiation related to M~51b is, and provide several maps of physical quantities. Various correlations and their discussion follow in Sect.~\ref{sec:discussion}. Our main conclusions are summarised in Sect.~\ref{sec:conclusions}.

\section{Dataset} \label{sec:dataset}

\subsection{Multi-wavelength imagery data} \label{subsec:phot}

We acquired a total of 26 broadband images of M~51 from the DustPedia archive\footnote{\url{http://dustpedia.astro.noa.gr}} \citep{Davies_2017PASP..129d4102D, Clark_2018A&A...609A..37C}. This panchromatic dataset covers a broad wavelength range from the FUV to the FIR domain (see Fig.~\ref{fig:photometry_M51}), combining archival data from different ground-based and space telescopes: GALaxy Evolution eXplorer \citep[GALEX;][]{Morrissey_2007ApJS..173..682M}, Sloan Digital Sky Survey \citep[SDSS;][]{York_2000AJ....120.1579Y, Eisenstein_2011AJ....142...72E}, 2 Micron All-Sky Survey \citep[2MASS;][]{Skrutskie_2006AJ....131.1163S}, Wide-field Infrared Survey Explorer \citep[WISE;][]{Wright_2010AJ....140.1868W}, \textit{Spitzer} \citep{Werner_2004ApJS..154....1W}; and \textit{Herschel} \citep{Pilbratt_2010A&A...518L...1P}. The retrieved images are essential in our modelling, since they will be used to construct the model and will be directly compared to the simulated output images. An H$\alpha$ image of M~51 was downloaded from the SINGS \citep{Kennicutt_2003PASP..115..928K} database\footnote{\url{https://irsa.ipac.caltech.edu/data/SPITZER/SINGS}}. The system was observed with the 2.1~m KPNO telescope, using a 2K~$\times$~2K CCD which has a field-of-view (FOV) of 10$^{\arcmin}$, and a pixel scale of 0.305${\arcsec}$. The narrow-band filter at 6573~\AA \, was used, with an exposure time of 900~seconds. The stellar continuum was already subtracted from the image, with foreground stars being identified as negative point-like residual features. We removed those features by interpolating over the data. The final image has a photometric accuracy within 15\%.

\subsection{Foreground stars identification} \label{subsec:prep}

All images were pre-processed with a semi-automatic pipeline as was developed in \textsc{PTS}\footnote{\url{http://www.skirt.ugent.be/pts8/_p_t_s.html}} \citep[Python Toolkit for \textsc{SKIRT}\footnote{\url{http://www.skirt.ugent.be/root/_landing.html}};][]{Verstocken_2020A&A...637A..24V}. First, we identify and exclude foreground stars in our images by searching the 2MASS All-Sky Catalog of Point Sources \citep{Cutri_2003yCat.2246....0C}. If the queried point sources have a local peak above a 3-sigma significance level, then they are identified as foreground stars, and they are subtracted together with their neighbouring pixels by the local background. The remaining features from the source subtraction are interpolated over the data, complemented with random noise. A manual inspection is performed afterwards for every image, to ensure that none of the regions within the galaxy were identified as point sources by mistake. We applied this procedure for the GALEX, SDSS, H$\alpha$, 2MASS, WISE, and \textit{Spitzer} images, since the contribution from foreground stars in the \textit{Herschel} bands is negligible.

\subsection{Background subtraction and Galactic extinction} \label{subsec:skysub}

The next step in the pipeline is to correct the images for background emission and Galactic extinction. We define an elliptical region that encircles both galaxies and is centred on M~51a (seen as the green ellipse in Fig.~\ref{fig:photometry_M51}). In order to estimate the large-scale variation of pixel values around M~51, we create a low-resolution background map with \textsc{Photutils} \citep{Bradley_Photutils}, dividing each image into a mesh-grid of rectangular regions. Each region is a square box with a side of $6 \times \text{FWHM}$ of the particular image. Of course, all pixels inside the green ellipse, associated with M~51 are masked. The final background map and its standard deviation are then generated by interpolation of the low-resolution map within the central ellipse.

Finally, we correct all images that are affected by Galactic extinction. We obtain the \textit{V}-band attenuation, $A_\mathrm{V}$, from the IRSA Dust Extinction Service\footnote{\url{https://irsa.ipac.caltech.edu/applications/DUST}}, in order to determine the attenuation in the UV optical and near-infrared (NIR) filters. We assume an average extinction law in the Milky-Way (MW) with $R_\mathrm{V}=3.1$ \citep{Cardelli_1989ApJ...345..245C}.

\begin{table}
\caption{Region definitions for the M~51 interacting system.}
\begin{center}
\resizebox{\textwidth}{!}{
\begin{threeparttable}
\begin{tabular}{lccccc}
\hline 
\hline 
Galaxy & R.A. (J2000) & DEC. (J2000) & $a$      & $b$      & P.A.   \\
       & [deg]        & [deg]        & [arcsec] & [arcsec] & [deg]  \\
\hline
M~51a  & 202.4721    & $+47.175159$ & 261.1     & 225.7    & 309.1  \\
M~51b  & 202.4981    & $+47.264645$ & 100.4     & 76.14    &  14.3  \\
M~51   & 202.4697    & $+47.195199$ & 357.4     & 288.1    & 309.1  \\
\hline
\end{tabular}
\begin{tablenotes}
Where $a$ is the semimajor axis, and $b$ is the semiminor axis of a particular ellipse. P.A. is the position angle of the disc according to \citet{Sheth_2010PASP..122.1397S}.
\end{tablenotes}
\end{threeparttable}}
\label{tab:regions}
\end{center}
\end{table}

\subsection{Global photometry} \label{subsec:apphot}

We perform our own global photometry, to ensure consistency over the measured flux densities between observed and mock images of M~51. We drew three different elliptical regions, as they are depicted in Fig.~\ref{fig:photometry_M51}. We used a fixed aperture for all images and for each target case, while ensuring that the ellipses of M~51a and M~51b do not overlap. The process of generating the aperture ellipse of each target is described in \citet{Clark_2018A&A...609A..37C}. The characteristics of each region are given in Table~\ref{tab:regions}. We calculate the uncertainties by adding in quadrature the calibration uncertainty, and the mean variation from the background. The measured flux densities and uncertainties of the observed images are given in Table~\ref{ap:photometry}. We would like to point out that the images of MIPS~70- and 160~$\mu$m, as well as their respective fluxes are not used in our modelling since the PACS~70- and 160~$\mu$m images surpass them in spatial resolution. In total we use 25 images and the aperture photometry of M~51 (green ellipse) to constrain our radiation transfer model. The photometry of the individual galaxies can be considered as ancillary information, and it is only used to disentangle the IRAC~3.6$~\mu$m flux density of the two galaxy components of the interacting pair.

According to previous studies \citep[e.g.][]{Toomre_1972ApJ...178..623T, Salo_2000MNRAS.319..377S}, the relative distance between the two galaxies is of the kpc order (20-50~kpc). In the interest of modelling, we adopted a common distance of $\sim7.91\pm0.87~$Mpc \citep{Sheth_2010PASP..122.1397S}. The implications of this choice are discussed in Sect.~\ref{subsec:dist_impl}. In addition, we estimate the inclination angles of M~51a and M~51b to be $32.6~\deg$ and $40.5~\deg$ from a face-on view, respectively, based on the method presented in \citet{Mosenkov_2019A&A...622A.132M}. Finally, the position angles given in Table~\ref{tab:regions} are based on the values provided by \citet{Sheth_2010PASP..122.1397S}.

\subsection{SED fitting with \textsc{CIGALE}} \label{sec:cigale}

During the last decade, modelling the spectral energy distribution (SED) has become increasingly a commonplace, that offers information about the global properties of any given galaxy \citep[see][for an overview of the available SED fitting tools and techniques]{Walcher_2011Ap&SS.331....1W, Conroy_2013ARA&A..51..393C, Baes_2020IAUS..341...26B}. As a first attempt to extract the intrinsic properties of M~51 and its galaxy components, we chose to fit their measured integrated fluxes with the SED fitting code \textsc{CIGALE}\footnote{\url{https://cigale.lam.fr}} \citep{Noll_2009A&A...507.1793N, Boquien_2019A&A...622A.103B}. \textsc{CIGALE} is equipped with several modules, each one describing a different physical component. These individual modules can be combined to create a library of SED templates, which will be fitted to the data. From these modules, we use a flexible star-formation history (SFH) \citep{Ciesla_2016A&A...585A..43C}, \citet{Bruzual_2003MNRAS.344.1000B} simple stellar population (SSP) libraries, and a modified \citet{Calzetti_2000ApJ...533..682C} attenuation law. The \citet{Calzetti_2000ApJ...533..682C} starburst attenuation curve is combined with the \citet{Leitherer_2002ApJS..140..303L} curve between the Lyman break and 150~nm, to produce a flexible attenuation law that can be adapted on vastly different galaxies. For the dust emission we use the \textsc{THEMIS}\footnote{\url{https://www.ias.u-psud.fr/themis/THEMIS\_model.html}} \citep[The Heterogeneous Evolution Model for Interstellar Solids;][]{Jones_2013A&A...558A..62J, Jones_2017A&A...602A..46J, Kohler_2014A&A...565L...9K} dust model. The \textsc{THEMIS} framework was built upon the optical properties of amorphous hydrocarbon and amorphous silicate materials as have been measured in the laboratory, and is qualitatively consistent with many dust observables, including: the observed FUV-NIR extinction, and the shape of the IR to mm dust thermal emission. In order to fit the H$\alpha$ flux we used the low resolution data as described in \citet{Boquien_2019A&A...622A.103B}. For this study, we adopted exactly the same parameter grid as in \citet[][for details see their Table~1]{Nersesian_2019A&A...624A..80N}, with the only difference being that we used a \citet{Chabrier_2003PASP..115..763C} IMF instead of a \citet{Salpeter_1955ApJ...121..161S}, to be consistent with our radiative transfer modelling. The generated library includes over 80 million template SEDs. Finally, \textsc{CIGALE} adds in quadrature an extra 10\% of the observed flux as model error.

\begin{figure}[t!]
\centering
\includegraphics[width=\textwidth]{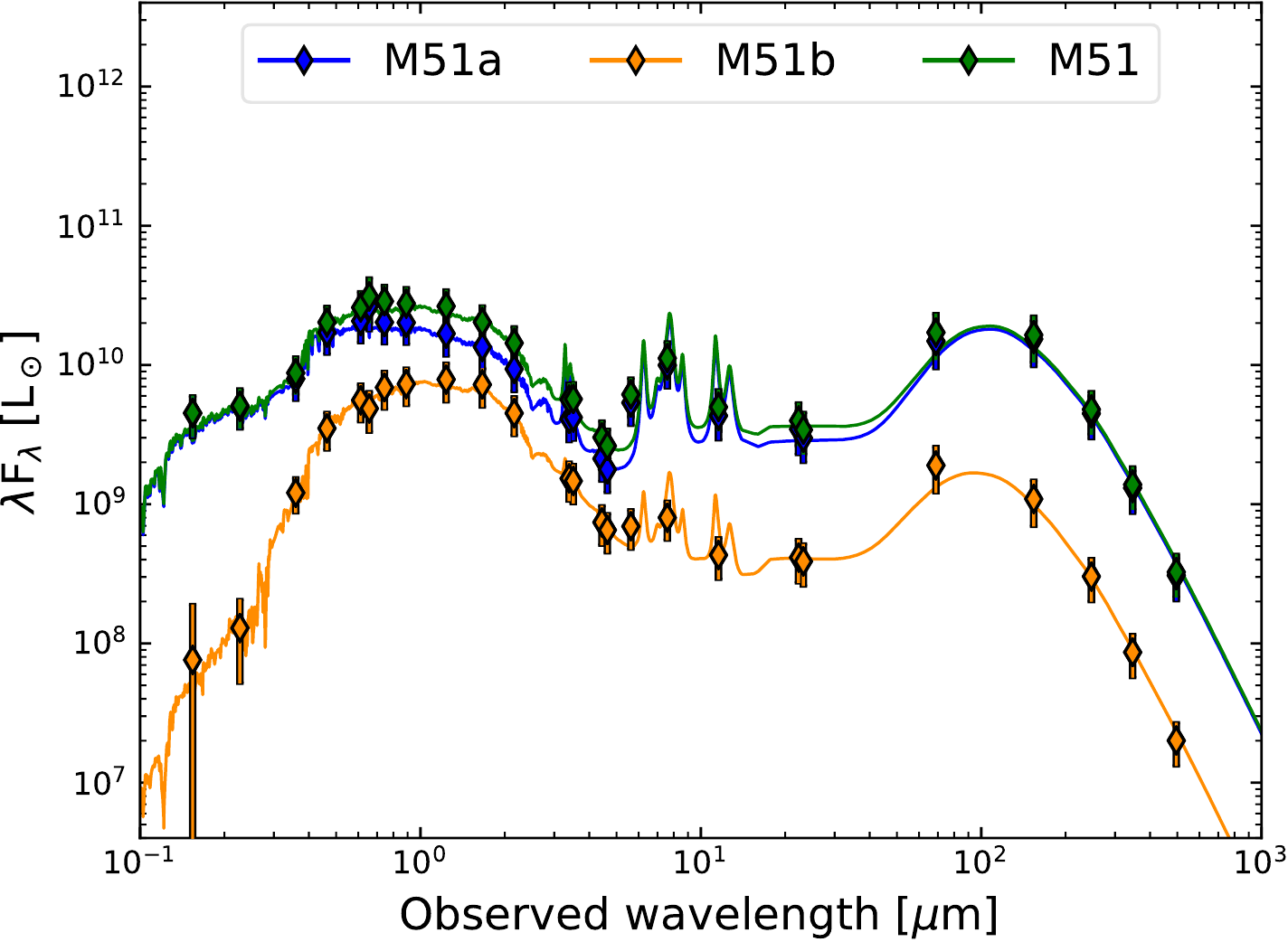}
\caption{SED of the M~51 interacting pair with \textsc{CIGALE}. The green line is the best-fitting model of M~51. The blue and orange lines are the best-fitting models of M~51a and M~51b, respectively. The inferred SEDs match the luminosities as measured from the aperture photometry shown in Fig.~\ref{fig:photometry_M51}.}
\label{fig:csed}
\end{figure}

Figure~\ref{fig:csed} shows the resulting best SEDs for all elliptical regions as defined in Table~\ref{tab:regions}. The diamonds correspond to the 25 measured integrated luminosities for each elliptical region. The model SEDs match the observations closely and fall within the measured uncertainties. Comparing the SEDs of M~51a (blue data) and M~51b (orange data) in relation to the total emission of the system (green data), we notice that most, if not all, of the UV and dust emission in M~51 arises from M~51a. In fact, the peak of the dust luminosity in the FIR from M~51a is about one order of magnitude higher than the peak of dust luminosity in M~51b. On the other hand, the SED of M~51b is dominated by the optical and NIR emission of an older stellar population without any indication of strong, ongoing star-formation activity.

This is further confirmed through the inferred SFRs from the global SED fitting. The SFR of M~51b is about $0.026\pm0.023~\text{M}_\odot~\text{yr}^{-1}$. Interestingly enough, M~51a forms stars on a moderate rate ($3.48\pm0.35~\text{M}_\odot~\text{yr}^{-1}$) which does not differ from SFRs measured in isolated galaxies. The stellar mass of M~51a is estimated to be equal to $1.37\pm0.38\times10^{10}~\text{M}_\odot$, and of M~51b is $1.35\pm0.26\times10^{10}~\text{M}_\odot$. Therefore, M~51 classifies as a major merger. The SFR of M~51 is measured to be $3.30\pm0.41~\text{M}_\odot~\text{yr}^{-1}$, slightly lower than that of the subsystem M~51a, but still consistent within the error bars. A possible explanation is that by incorporating the optical emission of M~51b to the system, the ratio of UV-to-optical decreases which results in a lower SFR. The total dust mass of the system is approximately $3.40\pm0.65\times10^7~\text{M}_\odot$, while the attenuation in the FUV is $1.98\pm0.15~\text{mag}$. These quantities are representative of the interacting system and they will be used as normalisations of the various modelling components in our radiative transfer simulations.

\section{Modelling approach} \label{sec:3dmodel}

In order to build a 3D model for M~51a and its companion, we follow the same strategy that was first introduced in \citet{De_Looze_2014A&A...571A..69D}, and later revised, and refined by \citet{Verstocken_2020A&A...637A..24V}. We take advantage of the state-of-the-art radiative transfer code \textsc{SKIRT} \citep{Baes_2011ApJS..196...22B, Camps_2015A&C.....9...20C, Camps_2020A&C....3100381C}. \textsc{SKIRT} follows the Monte Carlo approach to simulate the physical processes of scattering, absorption, and thermal re-emission by dust, in different environments. \textsc{SKIRT} contains a large collection of geometries and geometry decorators \citep{Baes_2015A&C....12...33B}, grid structures \citep{Saftly_2013A&A...554A..10S, Saftly_2014A&A...561A..77S, Camps_2013A&A...560A..35C}, and hybrid parallelisation techniques \citep{Verstocken_2017A&C....20...16V}. In our model, we make use of two special classes in \textsc{SKIRT}: the \texttt{ReadFitsGeometry} and \texttt{OffsetGeometryDecorator}. The \texttt{ReadFitsGeometry} loads a given 2D map into \textsc{SKIRT}, and converts it into a 3D disc component through image de-projection and by applying a perfect exponential profile in the vertical direction \citep[see][]{De_Looze_2014A&A...571A..69D}. The \texttt{OffsetGeometryDecorator} adds an offset component or a 2D map, in our case the inclusion of the companion galaxy M~51b, to the main geometry of M~51a \citep{Camps_2015A&C.....9...20C, Camps_2020A&C....3100381C}.

\begin{figure*}[!t]
\centering
\includegraphics[width=16cm]{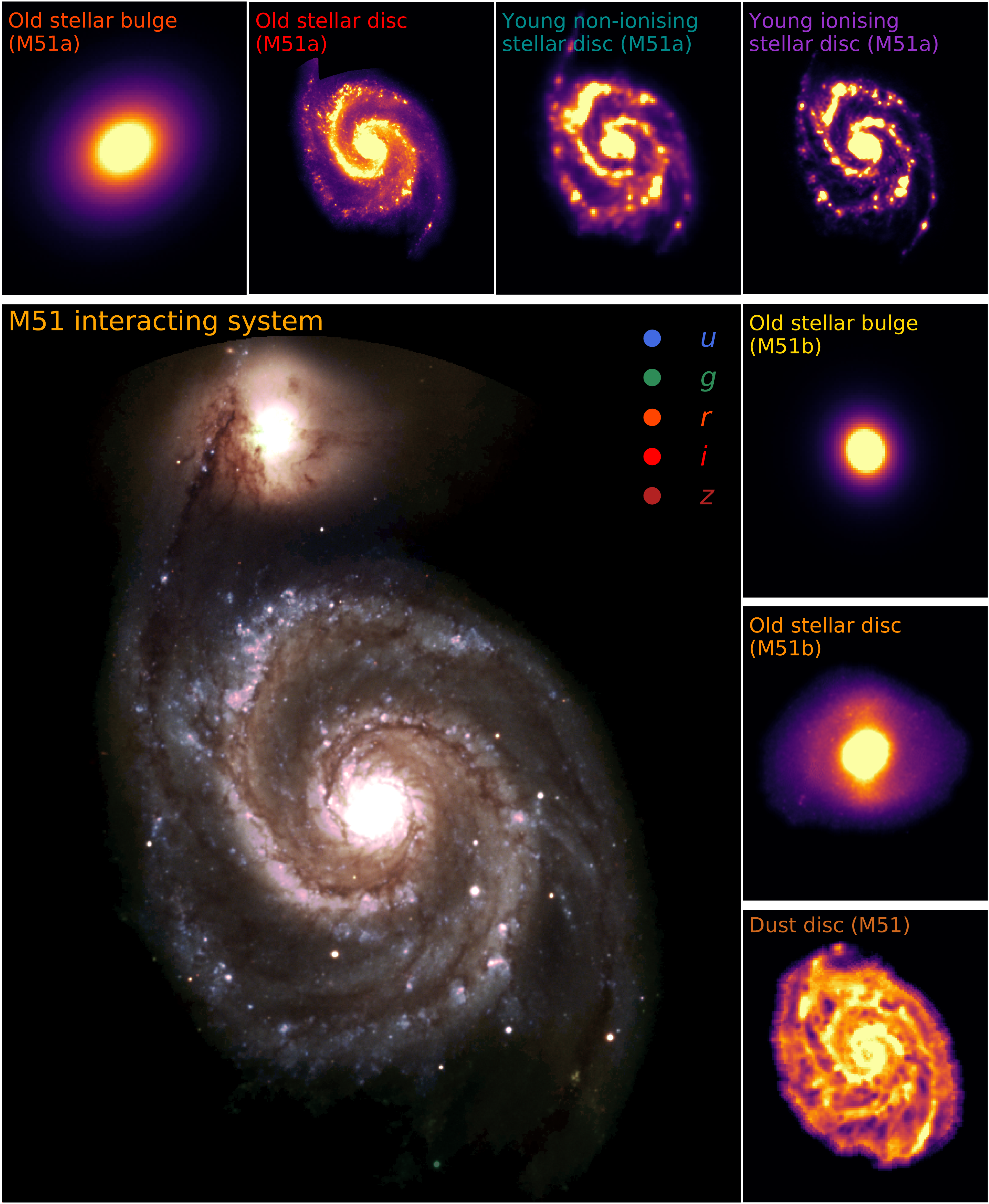}
\caption{The M~51 interacting system and its components. On the bottom left panel an optical image of M~51 is shown, created by combining the SDSS~$u, g, r, i, z$ images. On the top row and right column, a 2D representation of the various model components is depicted. From left to right: a zoomed-in view of M~51a's old stellar bulge, M~51a's old stellar disc, young non-ionising stellar disc, and young ionising stellar disc. From top to bottom: a zoomed-in view of M~51b's old stellar bulge, old stellar disc, and the combined dust distribution of M~51a and M~51b.}
\label{fig:comp_maps_m51}
\end{figure*}

In general, our model treats both galaxies as a uniform system with multiple components. Table~\ref{tab:model_params} gives an overview of the parameters of each input geometry, while Fig.~\ref{fig:comp_maps_m51} illustrates an optical image of M~51 with its various stellar and dust spatial distributions. Specifically for M~51a, we assume four different stellar components and a thin disc of dust based on the standard geometric model of spiral galaxies \citep{Xilouris_1999A&A...344..868X, Popescu_2000A&A...362..138P}. The novelty of this study is that we complement our model with an extra stellar component, M~51b. A common dust grid was generated, that contains the dusty discs of both galaxies. The inclusion of M~51b in the model will provide new insight on the interaction of the radiation field of a pre-merger, and will allow us to quantify, for the first time, the percentage of radiation from a neighbouring galaxy that goes into the dust heating of the main galaxy, and vice versa. In the following sections, we discuss briefly the different modelling components and highlight where we deviate from the parameterisation of \citet{De_Looze_2014A&A...571A..69D}. For a more complete description of our methodology we refer the reader to \citet{Verstocken_2020A&A...637A..24V}.

\subsection{Stellar components of M~51a}

We used observational images at various wavelengths to describe the geometrical distribution of the different stellar components in M~51a. We considered four stellar components, each one associated with a template SED and a 3D spatial geometry. The ages of each stellar component are in line with \citet{De_Looze_2014A&A...571A..69D}, however in our analysis we use a \citet{Chabrier_2003PASP..115..763C} initial mass function (IMF) instead of a \citet{Kroupa_2002Sci...295...82K} IMF, and the \citet{Bruzual_2003MNRAS.344.1000B} SSP templates instead of the \citet{Maraston_2005MNRAS.362..799M} SSP library. 

The old stellar population is described by two components: an old stellar bulge and a disc. We used the \textit{Spitzer}~IRAC~3.6~$\mu$m map, indicative of the older stars in the galaxy, to constrain this particular stellar population. However, the emission of the bulge and disk is entangled. Thus, a bulge-to-disc decomposition needs to be performed in order to retrieve the distribution of the old stellar population in the disc. The bulge is modelled with a flattened S\'{e}rsic profile and a fixed luminosity. The decomposition parameters of the S\'{e}rsic geometry were taken from the S$^4$G database\footnote{\url{https://www.oulu.fi/astronomy/S4G_PIPELINE4/MAIN}} \citep[\textit{Spitzer} Survey of Stellar Structure in Galaxies:][]{Sheth_2010PASP..122.1397S, Salo_2015ApJS..219....4S}. Then, the old stellar population disc is obtained by subtracting the bulge from the total observed 3.6~$\mu$m emission of M~51a. For this procedure it is assumed that the contamination from dust emission in the central region is negligible. The typical age of this stellar population is assumed to be $\sim 10$~Gyr with a fixed solar metallicity $Z = 0.02$ \citep[as reported by][]{Bresolin_2004ApJ...615..228B, Moustakas_2010ApJS..190..233M, Mentuch_Cooper_2012ApJ...755..165M}. The pixel scale resolution of the old stellar map is $0.75\arcsec$ (or 28.75~pc at the distance of M~51).

The third stellar component is the young non-ionising stellar population (yni) in the disc, with an assumed age of 100~Myr and a fixed solar metallicity $Z = 0.02$. The GALEX~FUV image is best suited to constrain this stellar component, since young stars (unobscured by dust) dominate this specific spectral range. We corrected the GALEX~FUV map for dust attenuation by generating a FUV attenuation map based on the prescriptions of \citet{Cortese_2008MNRAS.386.1157C}. These authors derive the FUV attenuation based on a polynomial function of the total infrared (TIR)-to-FUV ratio, on spatially resolved scales. The coefficients of the polynomial are determined for different values of the observed FUV$-r$ colour. The TIR emission map is constructed from the MIPS~24~$\mu$m, PACS~70~$\mu$m, and PACS~160~$\mu$m images, following the prescriptions described in \citet{Galametz_2013MNRAS.431.1956G}. Consequently, after we convolve the FUV, FUV$-r$, and TIR maps to the PACS~160~$\mu$m resolution (i.e. $4.0\arcsec$ or 153.4~pc at the distance of M~51), we were able to retrieve the intrinsic FUV emission of the young non-ionising stellar population. Inspecting this map closely we notice that the stellar bridge that connects the two galaxies is most prominent here. 

The fourth and final stellar component is the young ionising stellar population (yi), tracing the ongoing, obscured star formation. For this component we adopted the SED templates from MAPPINGS III \citep{Groves_2008ApJS..176..438G} assuming an age of 10~Myr. To constrain its emission we combined the \textit{Spitzer}~MIPS~24~$\mu$m image with the continuum-subtracted H$\alpha$ map, relying on the prescription derived by \citet{Calzetti_2007ApJ...666..870C}. There are five parameters that define the MAPPINGS III templates, namely: the mean cluster mass ($M_\mathrm{cl}$), the gas metallicity ($Z$), the compactness of the clusters ($C$), the pressure of the surrounding ISM ($P_0$), and the covering fraction of the molecular cloud photo-dissociation regions ($f_\mathrm{PDR}$). We adopted the following parameters as our default values and according to \citet{De_Looze_2014A&A...571A..69D}: $Z = 0.02$, $M_\mathrm{cl} = 10^5~\mathrm{M}_\odot$, $\log C = 5.5$, $P_0/k = 10^6~K~\mathrm{cm}^{-3}$, and $f_\mathrm{PDR} = 0.2$. The corresponding maps of both young stellar populations were corrected for the emission originating from old stars by subtracting a scaled version of the 3.6~$\mu$m image \citep[][and references therein]{Verstocken_2020A&A...637A..24V}. The pixel scale resolution of the young ionising stellar map is $1.5\arcsec$ (or 57.5~pc at the distance of M~51).

\begin{table}[t]
\caption{Overview of the modelling ingredients and normalisations for M~51. The normalisations are the initial values in our parameter space, as derived from the \textsc{CIGALE} SED fitting of M~51. Bulge parameters: $n$ is the S\'{e}rsic index, $q$ is the intrinsic flattening factor, and $R_\mathrm{e}$ is the effective radius.}
\begin{center}
\resizebox{\textwidth}{!}{
\begin{tabular}{llr}
\hline 
\hline 
Component & Parameter & Value \\
\hline
\hline
\multicolumn{3}{l}{\textbf{M~51a}} \\
\hline
\multicolumn{3}{c}{\textbf{Old stellar population}} \\
\hline
\multirow[t]{4}{*}{Bulge} & $n$ & 0.995 \\
 & $q$ & 0.817 \\
 & $R_\mathrm{e}$~[pc] & 621.1 \\
\textit{Normalisation} & $L_\mathrm{old,~3.6}^\mathrm{bulge}~[\mathrm{L}_\odot]$ & $0.54\times10^9$ \\
\hline
\multirow[t]{3}{*}{Disc} & 2D geometry & IRAC~3.6~$\mu$m \\
 & $h_\mathrm{disc,~z}$~[pc] & 390.8 \\
\textit{Normalisation} & $L_\mathrm{old,~3.6}^\mathrm{disc}~[\mathrm{L}_\odot]$ & $3.67\times10^9$ \\
\hline
\hline
\multicolumn{3}{c}{\textbf{Young non-ionising stellar population}} \\
\hline
\multirow[t]{3}{*}{Disc} & 2D geometry & GALEX~FUV \\
 & $h_\mathrm{disc,~z}$~[pc] & 195.4 \\
\textit{Normalisation} & $L_\mathrm{yni,~FUV}~[\mathrm{L}_\odot]$ & $2.15\times10^{10}$ \\
\hline
\hline
\multicolumn{3}{c}{\textbf{Young ionising stellar population}} \\
\hline
\multirow[t]{3}{*}{Disc} & 2D geometry & H$\alpha+0.031\times \mathrm{MIPS}~24~\mu$m \\
 & $h_\mathrm{disc,~z}$~[pc] & 97.7 \\
\textit{Normalisation} & $L_\mathrm{yi,~FUV}~[\mathrm{L}_\odot]$ & $6.61\times10^{9}$ \\
\hline
\hline
\multicolumn{3}{l}{\textbf{M~51b}} \\
\hline
\multicolumn{3}{c}{\textbf{Old stellar population}} \\
\hline
\multirow[t]{4}{*}{Bulge} & $n$ & 2.58 \\
 & $q$ & 0.564 \\
 & $R_\mathrm{e}$~[pc] & 222.4 \\
\textit{Normalisation} & $L_\mathrm{old~bulge,~3.6}^\mathrm{M51b}~[\mathrm{L}_\odot]$ & $0.37\times10^9$ \\
\hline
\multirow[t]{3}{*}{Disc} & 2D geometry & IRAC~3.6~$\mu$m \\
 & $h_\mathrm{disc,~z}$~[pc] & 370.9 \\
\textit{Normalisation} & $L_\mathrm{old~disc,~3.6}^\mathrm{M51b}~[\mathrm{L}_\odot]$ & $1.11\times10^9$ \\
 \hline
\hline
\multicolumn{3}{l}{\textbf{M~51}} \\
\hline
\multicolumn{3}{c}{\textbf{Dust}} \\
\hline
\multirow[t]{3}{*}{Disc} & 2D geometry & FUV attenuation map \\
 & $h_\mathrm{disc,~z}$~[pc] & 195.4 \\
\textit{Normalisation} & $M_\mathrm{dust}~[\mathrm{M}_\odot]$ & $3.40\times10^7$ \\
\hline
\end{tabular}}
\label{tab:model_params}
\end{center}
\end{table}

\subsection{Stellar component of M~51b }

The lack of major, ongoing star formation in M~51b, led us to assume that all of its stellar radiation stems from an older stellar population. In addition, the bulge of the galaxy is very bright in the optical and NIR wavelengths (see Fig.~\ref{fig:photometry_M51}), which could cause a strong elongated feature in the simulated images through the de-projection of the input map. In order to diminish the effect of this feature we replace the bulge with an analytical density profile. The bulge is modelled with a S\'{e}rsic profile, the parameters of which are given by the S$^4$G database \citep{Sheth_2010PASP..122.1397S, Salo_2015ApJS..219....4S}. Through a bulge-to-disc decomposition we subtract the emission of the bulge from the IRAC~3.6~$\mu$m image in order to get the stellar emission in the disc. The total luminosity is fixed such that it corresponds to the total IRAC~3.6~$\mu$m luminosity of M~51b. We assumed that the stellar age and metallicity of M~51b are similar to those of M~51a, that is $\sim 10$~Gyr with a fixed solar metallicity $Z = 0.02$ \citep[see also][]{Lee_2012ApJ...754...80L}, while we adopted a \citet{Chabrier_2003PASP..115..763C} IMF and the \citet{Bruzual_2003MNRAS.344.1000B} SSP templates. Finally, we assume that the overall contamination by the emission of aromatic features in the $3.6~\mu$m is negligible, since \citet{Boulade_1996A&A...315L..85B} found that their emission is concentrated on the central region of M~51b ($\le80~$pc). The pixel scale resolution of the old stellar component of M~51b is $0.75\arcsec$ (or 28.75~pc at the distance of M~51).

\subsection{Dust component of M~51}

The dust component is the combination of the thin dusty disc of M~51a and the dusty nucleus of M~51b. The diffuse dust disc is constrained through the FUV attenuation map, while the dust around the star-forming regions is modelled through a subgrid that relies on the MAPPINGS III SED templates. The advantage of using the FUV attenuation map is that we can trace the cold diffuse dust at higher spatial resolution, than if we had used the images at longer wavelengths (e.g. SPIRE~500~$\mu$m), which are less sensitive to the dust temperature and their resolution is significantly worse. Another difference between the modelling set up of \citet{De_Looze_2014A&A...571A..69D} and ours is that for the dust composition we used the DustPedia reference dust model \textsc{THEMIS} \citep{Jones_2013A&A...558A..62J, Jones_2017A&A...602A..46J, Kohler_2014A&A...565L...9K}, instead of a \citet{Draine_Li_2007ApJ...657..810D} dust mixture. The adopted \textsc{THEMIS} model is for the Milky-Way diffuse ISM, and we assume a constant fraction of aromatic hydrocarbons throughout the disc. As shown in the bottom right panel of Fig.~\ref{fig:comp_maps_m51}, the dust map highlights nicely the complex spiral structure of M~51a, but also reveals that any significant amount of dust in M~51b is concentrated in the vicinity of its central regions. The SDSS image also shows some strong dusty features along the tidal bridge and in the foreground of M~51b, however they are not resolved in the \textit{Herschel} bands. The pixel scale resolution of the dust map is $4.0\arcsec$ (or 153.4~pc at the distance of M~51).

\subsection{Including the 3D feature}

The use of the \texttt{ReadFitsGeometry} class in \textsc{SKIRT} allows the construction of complex 3D geometries from 2D maps, ensuring that the flux density is conserved during the conversion from 2D to 3D \citep{De_Looze_2014A&A...571A..69D}. To create the 3D distribution of the disc components, we assigned, to each one of them, an exponential profile of different scale heights, $h_\mathrm{z}$, based on previous estimates of the vertical extent of edge-on galaxies \citep{De_Geyter_2014MNRAS.441..869D}. Specifically, \citet{De_Geyter_2014MNRAS.441..869D} found a mean ratio of scale length to scale height, of 8.26. The scale lengths of M~51a and M~51b were obtained from the S$^4$G database, and subsequently we compute a scale height of 390.8~pc for the old stellar disc of M~51a, and 370.9~pc for the old stellar disc of M~51b. The young non-ionising and dust discs share the same scale height, that is, half of the scale height of the old stellar population of M~51a, while the scale height of the young ionising stellar disc is a quarter that of the old stellar disc of M~51a \citep{De_Looze_2014A&A...571A..69D, Viaene_2017A&A...599A..64V, Verstocken_2020A&A...637A..24V}. Each 2D map is de-projected on a face-on view (i.e. $i = 0~\deg$), and then it is smeared out in the vertical direction according to the given exponential profile. 

\subsection{Set up of the \textsc{SKIRT} simulations}

Due to the inherent geometrical complexity of the system, a few simplifying assumptions needed to be made. One such simplification was that M~51b has a common inclination angle as M~51a (i.e. $\sim33~\deg$). Another simplification was to assume that both galaxies are located in the same plane. Although, this assumption does not represent the current status of their relative distance separation, which according to the multiple encounter model is approximately 20~kpc, it certainly did 50-100~Myr ago during the passage of M~51b near M~51a \citep{Salo_2000MNRAS.319..377S}. We discuss the implications of this assumption to the contribution of dust heating by M~51b in Sect.~\ref{subsec:dist_impl}.

We generated a $k\text{-}d$ dust grid \citep{Saftly_2014A&A...561A..77S} based on the dust distribution of M~51, through which the photons of both galaxies will propagate in our simulations. The grid was made in a hierarchical way by splitting the spatial domain into 3D cells (voxels). The $x\times y\times z$ dimensions of the grid structure are $26.3\times26.3\times3.9~\text{kpc}^3$, containing approximately 1.4 million dust cells. The cell sizes were adjusted to the dust density distribution of M~51; grid cells are small where locations require high resolution, whereas cells can be much bigger elsewhere. The subdivision for individual cells stops when they contain a dust mass fraction below $8\times10^{-7}$. The size of a particular dust cell varies from 102~pc (maximum spatial resolution) to 3~kpc. Finally, we simulate datacubes of the observed radiation with a pixel scale of $4.0\arcsec$ or 153.4~pc.

\subsection{$\chi^2$ optimisation} \label{subsec:chi2_metric}

The normalisations assigned to each stellar and dust component in Table~\ref{tab:model_params} act as initial guess values and they were set as free parameters, determined through our optimisation procedure \citep{Verstocken_2020A&A...637A..24V}. The IRAC~3.6$~\mu$m luminosities of M~51a's old stellar bulge and M~51b, were kept fixed a priori. In total, we left four parameters in our model free, namely the FUV luminosity of the young non-ionising stellar disc ($L_\mathrm{yni,~FUV}$), the FUV luminosity of the young ionising stellar disc ($L_\mathrm{yi,~FUV}$), the IRAC~3.6~$\mu$m luminosity of the old stellar disc ($L_\mathrm{old,~3.6}^\mathrm{disc}$), and the total dust mass ($M_\mathrm{dust}$). The $L_\mathrm{yi,~FUV}$ was calculated by converting the SFR, derived from \textsc{CIGALE}, into a spectral luminosity. This was done by generating the \citet{Groves_2008ApJS..176..438G} spectrum and scaling it to the SFR. Then, the $L_\mathrm{yni,~FUV}$ was obtained by correcting the observed FUV luminosity with the FUV attenuation value retrieved by \textsc{CIGALE}, and then subtracting the found $L_\mathrm{yi,~FUV}$.

We would like to note that the $L_\mathrm{old,~3.6}^\mathrm{disc}$ was set as a free parameter to disentangle the light contamination by M~51b, as well as the possibility of dust contamination by the 3.3$~\mu$m aromatic feature. Global luminosities are distributed on the voxels according to the density distributions as prescribed by the physical maps. In order to determine the best model from our radiative transfer simulations, we follow a two-step minimisation procedure. The optimisation scheme was constructed in such a way so that it can reduce the computational cost of our simulations. 

The first phase of the optimisation, is to explore the parameter space around the initial guess values. To do so, we define a broad grid in the 4D parameter space of $(L_\mathrm{yni,~FUV},~L_\mathrm{yi,~FUV},~L_\mathrm{old,~3.6}^\mathrm{disc},~M_\mathrm{dust})$. For each free parameter we consider 5 grid points totalling to 625 simulated SEDs. Furthermore, we adopt a low-resolution spectrum of 115 wavelengths (0.1-1000$~\mu$m), which includes the effective wavelength of each band in our dataset for a direct comparison to the observed flux densities. At this stage, we do not require the spectral convolution of the simulated fluxes and images to the filter response curves. To ensure a satisfactory reconstruction of the global SEDs, we sample the emitted light in each wavelength with a sufficient number of photon packages ($10^6$). Moreover, we define six basic wavelength regimes: UV, optical, NIR, MIR, FIR, and submm, in order to assign a specific weight to each broadband filter, such that each wavelength regime is of equal importance \citep{Viaene_2017A&A...599A..64V, Verstocken_2020A&A...637A..24V}. The best-fitting model is determined through the comparison of the global observed and modelled SEDs: 

\begin{equation}
\chi^2 = \sum_X {w}_X 
\left( \frac{F_X^\text{obs} - F_X^\text{sim}}{\sigma_X^\text{obs}} \right)^2 \, ,
\label{eq:chi2global}
\end{equation}

\noindent where ${w}_X$, $F_X^\text{obs}$, $F_X^\text{sim}$, and $\sigma_X^\text{obs}$ are the assigned weight factor, global observed flux density, mock flux density, and observed error corresponding to waveband $X$, respectively. Then, we use the derived best-fitting values of this first exploratory stage, to narrow down the possible ranges of our 4D parameter grid, and to generate a new, refined, parameter grid space for the second round of simulations. 

For the second stage of the optimisation procedure, we define a high-resolution spectrum of 222 wavelength points, and shoot $5\times10^6$ photon packages per wavelength to ensure a realistic representation of the physical processes of emission, absorption, and scattering in M~51. Again, we use 5 grid points for each free parameter (625 simulated SEDs), while the option of spectral convolution is now enabled. The output of each simulation includes the SED of M~51 and a set of broadband images that can directly be compared to the observed images. In contrast to \citet{De_Looze_2014A&A...571A..69D}, who optimised only the global fluxes, we compute the $\chi^2$ metric based on the pixel-by-pixel difference of observed versus mock images in each band:

\begin{equation}
\chi^2 = \sum_X \sum_{p} 
{w}_X 
\left[\frac{\mu_{X}^\text{obs}(p) - \mu_{X}^\text{sim}(p)}{\sigma_{X}^\text{obs}(p)} \right]^2 \, ,
\label{eq:chi2local}
\end{equation}

\noindent where the first summation covers all filters $X$ for which data are available, the second sum loops over all pixels $p$ in the image corresponding to band $X$. The quantities $\mu_{X}^\text{obs}(p)$, $\mu_{X}^\text{sim}(p)$, and $\sigma_{X}^\text{obs}(p)$ are the observed surface density, mock surface density, and observed error in pixel $p$ of band $X$, respectively. Finally, ${w}_X$ is again the assigned weight factor given to the band $X$. 

We ran our simulations on the high-performance cluster of Ghent University. Each simulation of the first batch was evaluated under 50~min (on average) on a 12 CPU node. For the second batch of high-resolution simulations, the average time needed for each simulation to finish was about 4~h, again, on a 12 CPU node.

\section{Model results} \label{sec:results}

In the section that follows, we present the radiative transfer model that was best-fitted to the panchromatic dataset of M~51. We show the best-fitting SED along with the SEDs of the different stellar components, and we validate our results through an image comparison between the observed and the simulated images of M~51. 

\subsection{SED fitting with \textsc{SKIRT}} \label{subsec:ssed}

Once the best-fitting model is determined from our optimisation procedure, we run another simulation of that model but with twice the amount of photon packages (i.e. $10^7$ photon packages) per wavelength (222 wavelengths in total). This is done in order to reduce the inherent Monte Carlo noise and to generate high-quality images of M~51. In addition, we set individual simulations for each stellar component using the same dust grid configuration. This enables us to measure the pure fraction of radiation that is absorbed and/or scattered by the dust cells for each stellar source, and to examine their effective dust heating range. 

The radiative transfer model with the lowest local $\chi^2$ has 

\begin{equation}\label{eq:best_fitting_values}
\begin{cases}
\begin{split}
        L_\mathrm{old,~3.6}^\mathrm{disc} & = (2.08 \pm 0.43) \times 10^9~\text{L}_{\odot} \\
        L_\mathrm{yni,~FUV} & = (1.51 \pm 0.37) \times 10^{10}~\text{L}_{\odot} \\
        L_\mathrm{yi,~FUV}  & = (0.64 \pm 0.29) \times 10^{9}~\text{L}_{\odot} \\
        M_\mathrm{dust}     & = (5.09 \pm 0.83) \times 10^7~\text{M}_{\odot} \, .
\end{split}
\end{cases}
\end{equation}

\noindent The uncertainties on the parameter values were estimated by their probability distribution functions. In general, these parameters are in a good agreement with the initial guess values as derived from the 1D SED fitting with \textsc{CIGALE} (see normalisations in Table~\ref{tab:model_params}). Specifically, the most probable luminosities by the radiative transfer model are below the initial guess, with the most deviant case being for $L_\mathrm{yi,~FUV}$. The model luminosity of the young ionising stellar population is lower by 1.0~dex. On the other hand, the model predicts a dust mass that is approximately 1.5 times larger than the one derived by \textsc{CIGALE}. It appears that the differences in $L_\mathrm{yi,~FUV}$ and $M_\mathrm{dust}$ are linked. The model prefers adding more diffuse dust over a higher ionising radiation, required to warm up the dust in PDRs.

\begin{figure}[t!]
\centering
\includegraphics[width=\textwidth]{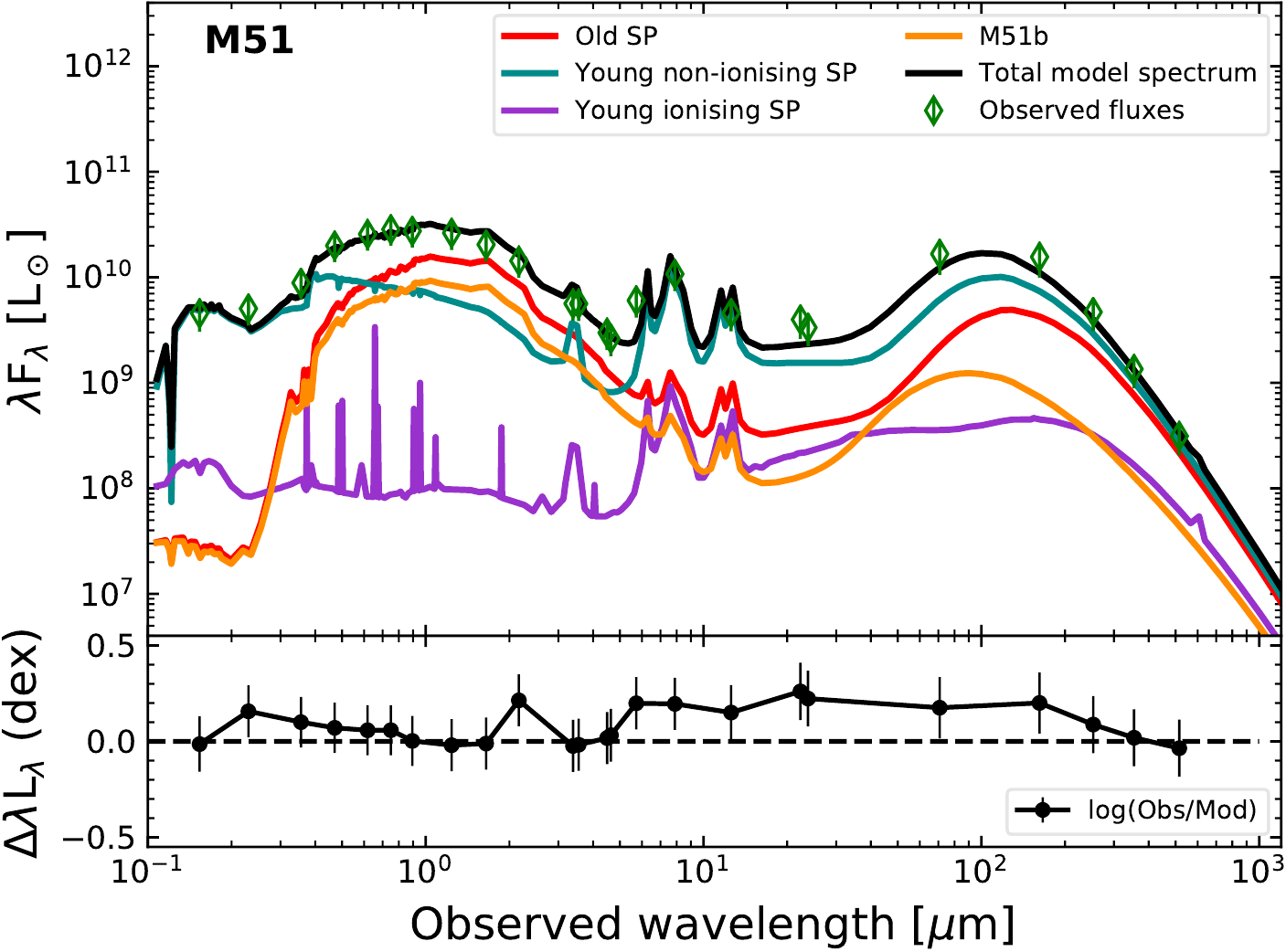}
\caption{\textit{Top panel}: Panchromatic SED of M~51. The black line is the best-fitting radiative transfer model, run at high-resolution. The green diamonds are the observed broadband luminosities of M~51 (see Table~\ref{tab:phot}). The red, cyan, and violet lines represent the SEDs for simulations with only the stellar components of M~51a: old, young non-ionising, and young ionising stellar population, respectively. The orange line represents the SED of M~51b. The dust component, which includes the dusty discs of both galaxies, is still present in these simulations. \textit{Bottom panel}: Difference in dex values between the observations and the mock luminosities.}
\label{fig:sed}
\end{figure}

In Fig.~\ref{fig:sed} we present the SED of the best-fitting model for M~51 along with its observed counterpart. The different coloured lines correspond to the SEDs of the various stellar components of M~51a, as well as the contribution of M~51b to the total SED. Due to the non-linear dependence among the absorbed energy and thermal re-emission by dust, summing up the individual SEDs does not result in the total model SED (black line). Furthermore, the dust emission of each SED component corresponds to the total absorbed radiation, re-processed by dust, in both galaxies. Overall, the comparison of the observations and the mock luminosities reveals an exceptionally good agreement between the two, with absolute differences not exceeding 0.26~dex. 

In detail, our model does a fine job fitting the UV, optical, NIR, and MIR (up to 4.6~$\mu$m) regions of the spectrum, with absolute differences below 0.1~dex. However, we note one significant deviation in the NUV band (0.23~$\mu$m), with the model underestimating the NUV luminosity by 0.16~dex. A similar deviation was observed in the radiative transfer model of M~51a by \citet{De_Looze_2014A&A...571A..69D}, but with a higher difference of about 0.3~dex. The NUV band still remains one of the hardest bands to model due to its dependence on the various components, and because the effects of dust extinction are more pronounced at this wavelength range \citep{Decleir_2019MNRAS.486..743D}. Another major discrepancy is seen for the 2MASS~$K_s$ band (2.16~$\mu$m). Surprisingly, our model underestimates the luminosity by 0.21~dex, despite the fact that it perfectly matches the emission in the 2MASS $J$ and $H$ wavebands (less than -0.01~dex). Recalling the imagery data of Fig.~\ref{fig:photometry_M51}, it is evident that the 2MASS images are less sensitive to the faint stellar emission. Consequently, the model struggles to properly sample the diffuse stellar emission on this particular wavelength range, resulting in the lower luminosity for the 2MASS~$K_s$ band. This difference could also be related to thermally pulsating asymptotic giant branch (TP-AGB) stars which have a dominant contribution at those wavelengths, and whose evolution is not well understood. Finally, the mismatch between model and observations could also be related to the specific choice of stellar libraries (and their inability to model the TP-AGB evolution correctly).

\begin{figure*}[t!]
\centering
\includegraphics[width=\textwidth]{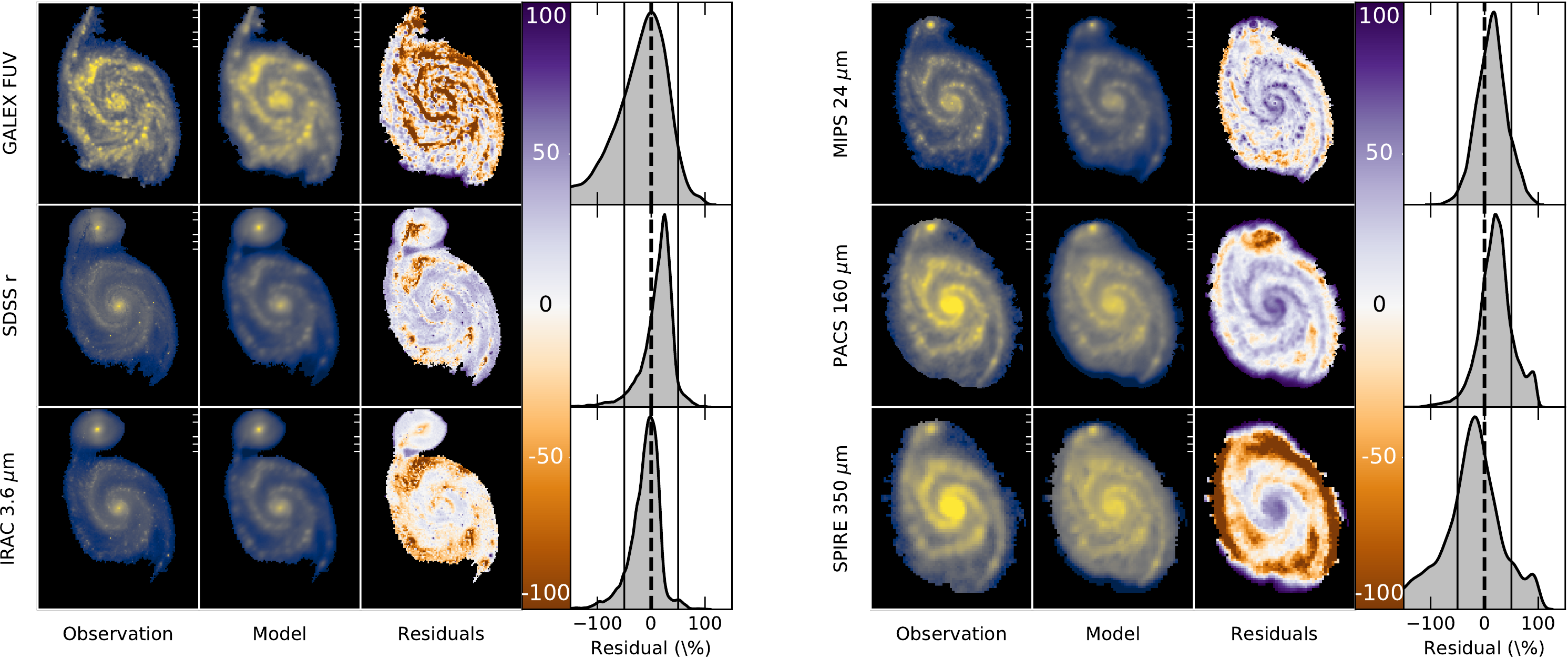}
\caption{Comparison of the simulated images with observations in selected wavebands for M~51. \textit{First column}: observed images, \textit{second column}, simulated images, \textit{third column}: residuals maps between observed and simulated images ($F_\mathrm{\nu,~obs} - F_\mathrm{\nu,~mod}/F_\mathrm{\nu,~obs}$). Positive values (in blue) mean that the model underestimates the observed emission, and negative values (in red) mean that the model overestimates the observations. \textit{Last column}: KDE distributions of the residual pixel values. The simulated images have the same pixel mask as the observed images. The colour coding of the first two columns reflects a normalised flux density. The selected wavebands are: GALEX~FUV, SDSS~$r$, IRAC~$3.6~\mu$m, MIPS~$24~\mu$m, PACS~$160~\mu$m, and SPIRE~$350~\mu$m. The vertical lines indicate the 0 (dashed line), and $\pm50$ (solid lines) percentage levels of the residuals.}
\label{fig:res_maps_m51}
\end{figure*}

Beyond 4.6~$\mu$m our model underestimates the emission of the aromatic feature carriers observed in the IRAC~I3~(5.8~$\mu$m), IRAC~I4~(8~$\mu$m), and WISE~W3~(12~$\mu$m), as well as the continuum emission in the 22-24~$\mu$m. The largest deviation in our model is for the WISE~W4~(22~$\mu$m) emission, with a difference of 0.26~dex. The model of M~51a by \citet{De_Looze_2014A&A...571A..69D}, similarly underestimated the dust emission at these wavelengths. The cause of this discrepancy appears to be associated with our assumption of a constant fraction of hydrocarbons throughout the system, as it is shown that the fraction of hydrocarbons varies within galaxies and from galaxy to galaxy \citep{Galliano_2008ApJ...679..310G, Peeters_2017ApJ...836..198P}. In low density regions, dust grains are prone to destruction by the intense stellar radiation field \citep[e.g.][]{Draine_2007ApJ...663..866D, Bendo_2008MNRAS.389..629B, Galliano_2018ARA&A..56..673G}, whereas in high density regions dust carriers may grow onto bigger grains through accretion and coagulation \citep[e.g.][]{Hirashita_2011MNRAS.416.1340H, Kohler_2012A&A...548A..61K, Kohler_2015A&A...579A..15K}. Unfortunately, these processes cannot be taken into account, as they have not been incorporated in \textsc{SKIRT} yet. Although, the framework to implement them is now ready \citep{Camps_2020A&C....3100381C}. Finally, the model, falls short to fit the peak of the dust emission in the FIR with the difference in PACS~70, and 160~$\mu$m being 0.18~dex and 0.2~dex, respectively, while the luminosities of the longest FIR wavelengths ($\ge 250~\mu$m) are in excellent agreement, with residuals below 0.1~dex. As we already mentioned, the small offset between the observed and model dust emission peak indicates that more UV radiation was needed to warm up the dust. 

Looking now at the individual component SEDs, it is evident that the young non-ionising stellar population (cyan line), of an assumed age of 100~Myr, is the dominant UV source in M~51a. The young non-ionising stars are also responsible for the bright aromatic features observed at 3.3-, 6.2-, 8-, and 11.3~$\mu$m. According to the \textsc{THEMIS} dust model, these features are due to amorphous hydrocarbon solids \citep{Jones_2013A&A...558A..62J, Jones_2017A&A...602A..46J, Galliano_2018MNRAS.476.1445G}. The bulk of the dust emission in the FIR is also provided through heating by the same stellar population. Furthermore, the lack of a strong ionising stellar component in the disc (violet line) could explain the underestimation of the MIR and FIR dust emission. It can further explain the surprisingly low $L_\mathrm{yi,~FUV}$, since this parameter was used to constrain this particular SED. On the other hand, we see that M~51b provides a significant contribution to the optical and NIR emission in the system. The absorbed and/or scattered radiation from M~51b warms up its own dust, as well as a considerable fraction of the diffuse dust of M~51a to high dust temperatures. This can be clearly seen from the dust emission induced by the companion galaxy, the peak of which is shifted toward a shorter wavelength, indicative of a warmer dust component.

\subsection{Image comparison} \label{subsec:obs_vs_mod}

Since the simulated images were used in the optimisation procedure to determine the best-fitting model, it is relevant to present a spatial comparison between the model and the observed broadband images. The spectral convolution of the simulated datacube generates a set of 24 mock images, all clipped to the same signal-to-noise mask of their observed counterpart. Figure~\ref{fig:res_maps_m51} illustrates the spatial comparison for a few selected broadband images, characteristic of the different wavelength regimes (UV, optical, NIR, MIR, FIR, submm). The first column shows the observations, the second column the simulations, and the third column the residual images. The fourth column of Fig.~\ref{fig:res_maps_m51} shows the kernel density estimation (KDE) of the residual values. 

A visual inspection of the residual images presented here, reveals the excellent agreement between model and observations for M~51, with the majority of the model pixels within $\pm50\%$ of their observed counterpart. A substantial fraction of those residuals is caused by the soft blur that is visible in the model images, a relic of the de-projection procedure. In the FUV band the model image compares quite well with the observation, with the KDE distribution of the residuals peaking near 10\%. The model is able to correctly reproduce the diffuse FUV emission in the inter-arm regions of the disc, yet it overestimates the emission across the spiral arms with residuals lower than -50\%. \citet{De_Looze_2014A&A...571A..69D} also found a similar result in their model of M~51a. They explained these variations in their model as a result of the assumed constant age of the young stellar populations. In reality, we expect that spiral arms are populated by stellar clusters of different ages \citep[see also][]{Wei_2020arXiv200706231W}. In addition, the model overestimates the emission from the bulge of M~51b up to 90\%. 

A very good match between model and observation is seen also for the optical image in the SDSS~$r$ band. The model accurately reproduces the observed image, with a narrow distribution and a peak at residual values close to 23\%. Very few residuals are below 0\% and they are located in the spiral arm nearest to the companion galaxy and in the tidal bridge. Likewise, the IRAC~$3.6~\mu$m residual map has a narrow distribution with a peak slightly shifted below 0\%. The map itself shows very few residuals, remaining mostly close at the interaction region and partially in the inner regions of the spiral arms. The diffuse old stellar emission and the emission from the bulge of each galaxy are well reproduced. With respect to the IRAC~$3.6~\mu$m image produced by \citet[][]{De_Looze_2014A&A...571A..69D}, the diffuse stellar emission was underestimated by 30\%. Our improved spatial reconstruction of the IRAC~$3.6~\mu$m image makes us confident that the old stellar population, i.e. the stellar mass in the interacting system, is accurately described by our model.

Despite the considerable difference found in the integrated MIPS~24~$\mu$m luminosity (0.22~dex or 40.2\%), the model accurately describes the spatially resolved diffuse emission from warm dust in the disc and inter-arm regions of M~51a. The pixel values in the simulated image are underestimated by 14.5\%, with the spiral structure being quite prominent in the residual map. Specifically, the model fails to reproduce the sites of obscured star-formation across the spiral arms with the relative differences exceeding 50\%. The bright centre of M~51b is also heavily underestimated (up to 85\%), while the footprint of the MIPS~24~$\mu$m PSF adds up to the observed discrepancies. The FIR simulated images and observations are in good agreement, with residual values having a narrow distribution peaking within $\pm22\%$. In the case of PACS~$160~\mu$m waveband, the model underestimates the dust emission mainly in the spiral arms and the central region of M~51a, with the diffuse dust in the inter-arm regions accurately represented. Both MIPS~24~$\mu$m and PACS~$160~\mu$m images are better reproduced in the model of \citet{De_Looze_2014A&A...571A..69D}, where the peak of the residual distribution for these two bands is very close to 0\%. 

Finally, the SPIRE~$350~\mu$m emission in the disc, a tracer of the cold diffuse dust, is overestimated up to 22\%. The centre of M~51a is underestimated with the relative residuals rising up to 55\%, while the edges of the disc are being greatly overestimated. Arguably, our simplification of placing M~51a and M~51b at the same distance from the observer combined with the lower dust density around the edges of the disc, allowed the photons of the companion galaxy to propagate through those regions, warming up the colder dust component (see also Sect.~\ref{subsec:heating_maps} and Sect.~\ref{subsec:physical_maps}). The same explanation applies to the conspicuous feature in the PACS~$160~\mu$m residual map around the region where the two galaxies blend in.

In conclusion, despite small deviations, we are confident that our radiative transfer model of M~51 describes the observed data sufficiently. Compared to the model of M~51a by \citet{De_Looze_2014A&A...571A..69D}, our model does a better job reconstructing the short wavelength (UV, optical, NIR) images, while it deviates for the longer wavelengths (MIR and FIR). The differences between the two models appear to be connected to some extent with the chosen dust mixture properties \citep[\textsc{THEMIS} vs][]{Draine_Li_2007ApJ...657..810D}. For example, the total dust mass we obtained for M~51 is $\sim 34\%$ lower than the dust mass $M_\text{dust} = 7.7\times10^7~\text{M}_\odot$ reported by \citet{De_Looze_2014A&A...571A..69D} just for M~51a. This is expected, since the dust in \textsc{THEMIS} is more emissive than in the \citet{Draine_Li_2007ApJ...657..810D} model, having both a lower emissivity index $\beta$ and a higher absolute opacity $\kappa_0$ value \citep[e.g., Fig.~4 of][]{Galliano_2018ARA&A..56..673G}. The inclusion or not of the companion galaxy is an important factor as well. The introduction of this extra component in our model, certainly helped to better sample the diffuse stellar emission in the optical and NIR wavebands. On the other hand, it resulted in the significant overestimation of the cold diffuse dust emission around the edges of M~51a. The direct influx of photons, by M~51b, could explain the extra amount of dust emission. We keep in mind these caveats when analysing the dust heating in Sect.~\ref{subsec:heating_maps}.     

\begin{figure*}[t!]
\centering
\includegraphics[width=\textwidth]{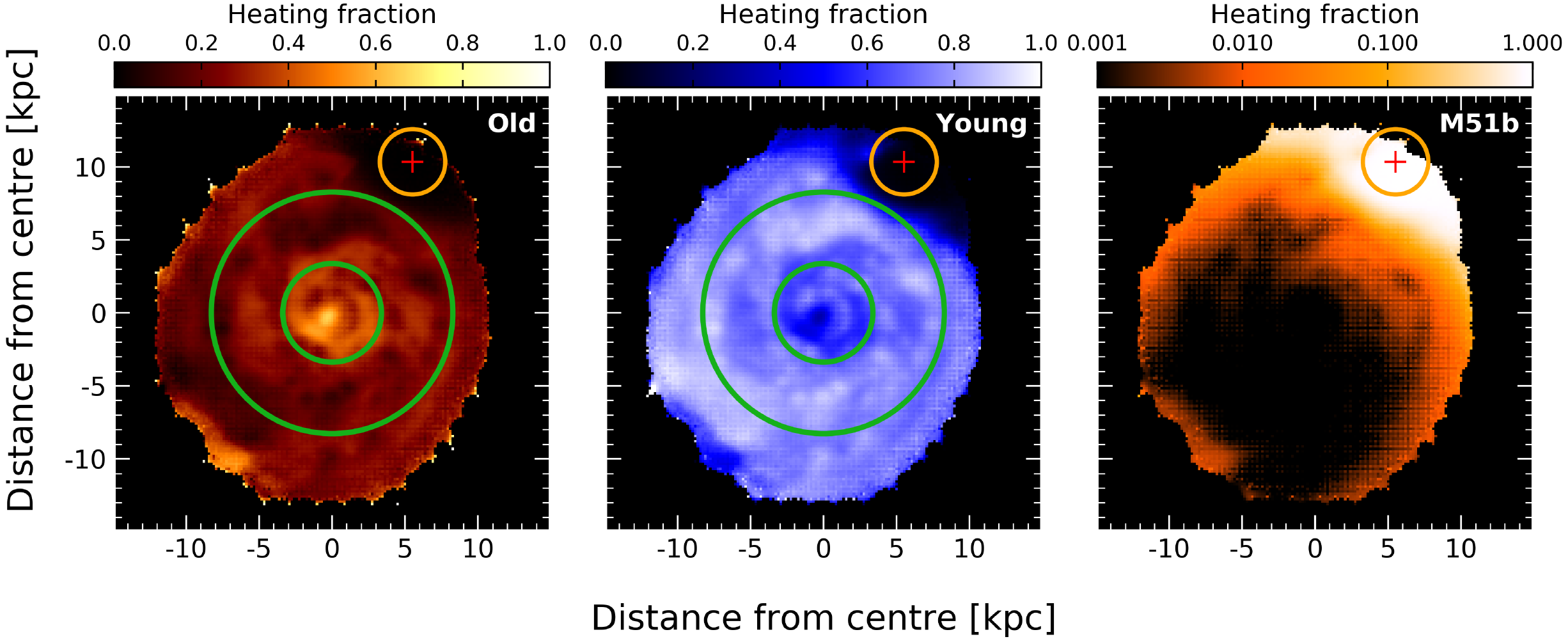}
\caption{Dust heating maps of a face-on view of M~51, as obtained from the 3D dust cell data. \textit{Left panel}: dust heating fraction by the old stellar population, \textit{middle panel}: dust heating fraction by the young stellar populations, and \textit{right panel}: dust heating fraction by M~51b. The rightmost map is shown on a $\log$-scale when the others are on a linear scale. The inner and outer green circles mark the 3~kpc and 8~kpc radii, respectively. The red cross marks the centre of M~51b. The gold circle indicates the extent of the dusty disc of M~51b at 2~kpc radius. All maps share a common pixel scale as the PACS~$160~\mu$m image, that is $4.0\arcsec$.}
\label{fig:m51_heating_maps}
\end{figure*}

\section{Dust heating results} \label{sec:dust_heating}

Cosmic dust has a major influence on how we observe galaxies, since it is responsible for the reddening and scattering of starlight. In fact, several studies have shown that dust can absorb roughly one third, in some cases even more than half, of the total stellar radiation in nearby, late-type galaxies \citep[e.g.][and references therein]{Viaene_2016A&A...586A..13V, Bianchi_2018A&A...620A.112B}. Based on the energy-balance principle, the stellar electromagnetic energy absorbed by dust is then re-processed and released in the form of thermal radiation at MIR, FIR, and submm wavelengths. We measure the global dust heating fraction simply by calculating the fraction of dust to bolometric luminosity: $f_\text{abs}^\text{SKIRT} = L_\text{dust}/\left(L_\text{stars} + L_\text{dust}\right)$. $L_\text{stars}$ is the total observed stellar emission and $L_\text{dust}$ is the total dust luminosity, computed by integrating the SED presented in Fig.~\ref{fig:sed}. For M~51, we find that 40.7\% of its total stellar radiation is absorbed by dust. A similar value of 41.2\% is retrieved from the global SED of M~51 as derived with \textsc{CIGALE} (see Fig.~\ref{fig:csed}). The advantage of a 3D radiative transfer model of M~51 over a simple 1D SED fitting is that we can disentangle the dust-heating fraction by the different stellar components not only on global scales but also on resolved, sub-kpc scales. Another advantage is that we can quantify the dust-heating fraction related to the companion galaxy, M~51b. In the following sections, we present the dust-heating fractions by the different stellar sources in our model, as well as the intrinsic properties and correlations for each galaxy in the interacting system.

\subsection{Heating maps} \label{subsec:heating_maps}

Each dust cell in our simulation holds information about the energy it absorbed and its origin. Subsequently, it is possible to disentangle the complex radiation field of the interacting system and quantify the dust-heating fraction from the different heating agents. We define the young heating fraction or $f_\text{young}$ as the fraction originating from the young non-ionising and young ionising stellar populations of M~51a. Likewise, we define the old heating fraction or $f_\text{old}$ as the heating fraction provided by the old stellar population of M~51a, and the heating fraction related to the absorbed radiation emitted by M~51b across the galactic disc of M~51, as $f_\text{M51b}$. We obtain the aforementioned heating fractions through:

\begin{equation} \label{eq:heating_fractions}
\begin{split}
    L^\text{abs}_\text{tot} & = L^\text{abs}_\text{old} + L^\text{abs}_\text{yni} + L^\text{abs}_\text{yi} +  L^\text{abs}_\text{M51b} \, , \\
    f_\text{old} & =  \frac{L^\text{abs}_\text{old}}{L^\text{abs}_\text{tot}} \, , \\
    f_\text{young} & =  \frac{L^\text{abs}_\text{yni} + L^\text{abs}_\text{yi}}{L^\text{abs}_\text{tot}} \, , \\
    f_\text{M51b} & = \frac{L^\text{abs}_\text{M51b}}{L^\text{abs}_\text{tot}} \, .
\end{split}
\end{equation}

\noindent where $L_\text{tot}^\text{abs}$ is the total stellar luminosity absorbed by dust for the interacting pair, $L_\text{yni}^\text{abs}$ and $L_\text{yi}^\text{abs}$ are the luminosities of the young non-ionising and young ionising stellar populations absorbed by dust, respectively, and $L^\text{abs}_\text{M51b}$ is the absorbed luminosity originally emitted by M~51b. 

Figure~\ref{fig:m51_heating_maps} shows the various dust-heating fractions from a face-on view of M~51. The left and middle panels show the dust-heating maps by the old and young stellar populations, respectively. The right panel depicts the dust-heating fraction provided by the older stellar population in M~51b. This particular heating map is in $\log$-scale in order to better visualise the heating fractions across the disc of M~51a. These maps are averaged along the line of sight. The inner and outer green circles mark the 3~kpc and 8~kpc radii, respectively, whereas the gold circle indicates the limits of the dusty disc of M~51b at 2~kpc radius. For the combined M~51 system, we find that on average, $5.8\%$ of the dust emission is attributed to the dust heating by the old stellar population of M~51b, of which $4.8\%$ is happening in the M~51a subsystem. $23\%$ of dust heating is supplied by the old stellar population of M~51a, while the remaining $71.2\%$ rises from the heating by the young stellar populations. Out of that $71.2\%$, $67.8\%$ springs from the young non-ionising stellar population, being the main heating agent, and $3.4\%$ comes from the young ionising stellar population. 

In order to have a fair comparison between our results and those by \citet{De_Looze_2014A&A...571A..69D}, we isolate the disc of M~51a and measure the respective heating fractions in a radius of 10~kpc from its centre. Thus, for M~51a we find a $f_\text{young}$ of $72.1\%$ and $f_\text{old}$ of $23.1\%$. These values are approximately 9\% higher and 14\% lower, respectively, than the values reported by \citet{De_Looze_2014A&A...571A..69D}. This difference can be linked back to the FUV luminosity which is used to constrain the emission by the young stellar populations, and the use of different extinction laws related to the corresponding, employed dust models. In the model of \citet{De_Looze_2014A&A...571A..69D}, the FUV emission is underestimated approximately by 18\% (see the residual distribution of the FUV map, in their Fig.~5). Since we selected our best-fitting model based on a local $\chi^2$ metric, the spatial distribution of the FUV emission in M~51a is more accurately reproduced, and thus it provides a more precise measurement of the young heating fraction in the disc. 

Inspecting the spatial variation of the young heating fractions in M~51a (middle panel of Fig.~\ref{fig:m51_heating_maps}), we see that for the most part, the young stellar populations are mostly located in the spiral arms (white coloured regions), and less in the inter-arm and central areas of the disc (regions of darker blue colours). On the other hand, paying a closer attention to the old heating map (left panel of Fig.~\ref{fig:m51_heating_maps}), we notice that the old stellar population dominates the radiation field at the very centre of the galaxy (below 0.5~kpc) with $f_\text{old}$ up to 68\%. However, we measure that the contribution of the young stellar population quickly rises to $57\%$ in the first kpc distance from the centre, and about $64\%$ in the inner 3~kpc radius. After the 3kpc distance, $f_\text{young}$ remains roughly constant, approximately $76\%$ within a radius of 8~kpc (see also Fig.~\ref{fig:rad_profiles} for the radial profiles of the heating fractions).

The rightmost panel of Fig.~\ref{fig:m51_heating_maps} shows the extent of the radiation field of M~51b. Most of the radiation that comes from the companion galaxy heats up the dust in the upper regions of M~51a, donating up to 38\% to that purpose. Besides, it is evident that photons can travel deep into the interstellar medium of M~51a, even making their way to the far side of the disc. However, it is easier for the photons to propagate in the less dense outskirts of M~51a's disc through scattering, than to pass through the more dense areas within the disc. Although $f_\text{M51b}$ remains under the percentage level around the outskirts of M~51a, it could partially explain the increased dust emission at the longest wavelengths that we discussed in Sect.~\ref{subsec:obs_vs_mod}. As for M~51b itself, it contains very little amount of dust, all concentrated in the central region of the galaxy and within a 2~kpc radius (gold circle). The dust, for the most part, is heated by the extremely strong, native stellar radiation field, up to $98\%$. The remaining 2\% is evenly contributed by the young and the old stellar populations of M~51a. This is no surprise, since it is less likely for M~51a photons to heat the dust of M~51b. First, because it is harder for photons to escape the dense ISM of M~51a, and secondly, because of the small extent of the dusty disc of M~51b.  

Notwithstanding that the two galaxies are in an early stage of their merging episode, and without any notable signs of their recent encounter, our findings showcase how the companion galaxy in a major merging system, such as M~51, could have a considerable effect on the energy budget of the main galaxy. It is also fair to assume that this effect could only be amplified in minor and major post-mergers. This could significantly impact the derived star-formation rates and ISM masses for such systems. Further investigation and quantification of the energy exchange between interacting pairs of different merging stages, through radiative transfer techniques, can be performed on cosmological simulations, like EAGLE \citep{Schaye_2015MNRAS.446..521S} or IllustrisTNG \citep{Nelson_2019MNRAS.490.3234N, Pillepich_2019MNRAS.490.3196P}.  

\subsection{Implications due to the relative distance separation} \label{subsec:dist_impl}

In this section we discuss how our choice to place M~51a and M~51b at the same distance may have affected the results that we obtain for the dust heating. In particular, the influence of the radiation field of M~51b on M~51a. By placing both galaxies in the same plane (centre-to-centre distance separation of $\sim10$~kpc), the contribution of heating by M~51b is maximised. As mentioned in Sect.~\ref{sec:intro}, M~51b can be located at a distance somewhere between 20~kpc \citep{Salo_2000MNRAS.319..377S} and 50~kpc \citep{Toomre_1972ApJ...178..623T} beyond the disc of M~51a. Although we tend to lean towards the most up-to-date reported distance (i.e. 20~kpc), here we investigate three different relative distances between the two galaxies, 20~kpc, 30~kpc, and 50~kpc. In order to measure the global $f_\mathrm{M51b}$ at these distances, we ran three additional radiative transfer simulations, containing just the contribution of M~51b and the dust distribution. The IRAC~3.6$~\mu$m luminosity of M~51b was kept fixed in all three cases ($1.48\times10^9~\mathrm{L}_\odot$). 

\begin{figure}[t!]
\centering
\includegraphics[width=\textwidth]{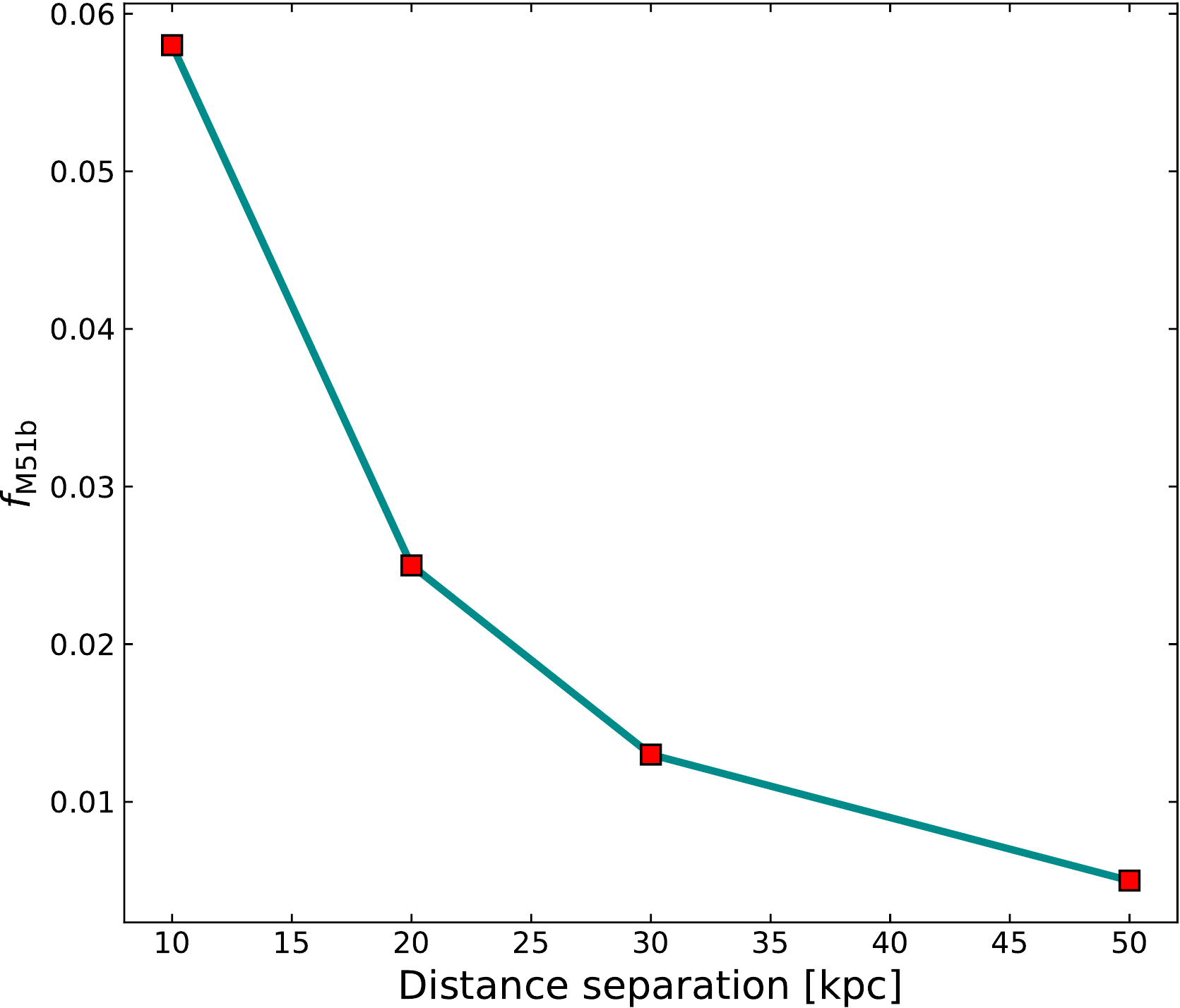}
\caption{Global dust heating fraction by M~51b against relative distance separation from the centre of M~51a.}
\label{fig:m51_distance_implications}
\end{figure}

Figure~\ref{fig:m51_distance_implications} shows the global dust heating fraction provided by M~51b to the system as a function of the relative distance separation between the two galaxies. As expected, the heating fraction $f_\mathrm{M51b}$ decreases with distance. Still, the dust heating by M~51b remains significant at a relative distance of 20~kpc, with $f_\mathrm{M51b}=2.5\%$. In fact, $f_\mathrm{M51b}$ is even higher for the cells in the inter-arm regions because they are more exposed to M~51b now. At a distance of 30~kpc $f_\mathrm{M51b}$ drops to 1.3\%, and remains below the percentage level at 50~kpc distance. Here we should stress the fact that the space in between M~51a and M~51b is not empty, but filled with tidal debris from their recent interaction \citep[e.g.][]{Watkins_2015ApJ...800L...3W}, and with hot gas that constitutes the intergalactic medium (IGM). The effect of the IGM on the radiation of M~51b is not accounted for in our simulations, since there is not yet an implementation for it in \textsc{SKIRT}. So, even if we have modelled these galaxies at their correct distances, it would have been hard to account for the IGM and its effect on the radiation.   

\begin{figure*}[t!]
\centering
\includegraphics[width=\textwidth]{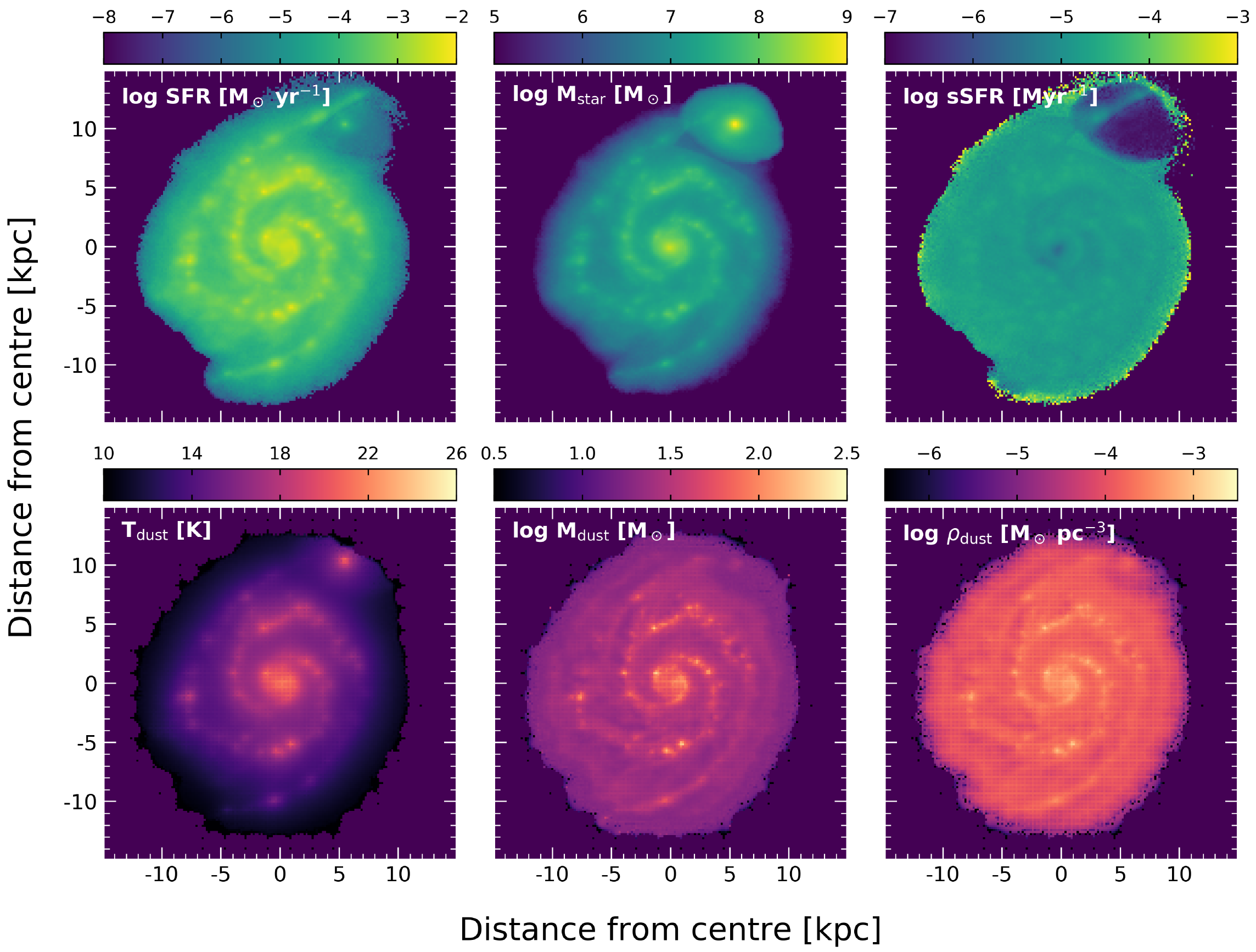}
\caption{Various physical maps of a face-on view of M~51. \textit{Top row}: from left to right, the SFR, stellar mass, and specific star-formation rate spatial maps. \textit{Bottom row}: from left to right, the dust temperature, dust mass, and the dust density spatial maps. All maps, except the dust temperature map, are in log-scale. All maps share a common pixel scale as the PACS~$160~\mu$m image, that is $4.0\arcsec$.}
\label{fig:m51_physical_maps}
\end{figure*}

\subsection{Maps of physical properties} \label{subsec:physical_maps}

In Fig.~\ref{fig:m51_physical_maps} we present a set of the most relevant physical maps for a face-on view of M~51, as approximated by our radiative transfer model. The information of each physical quantity depicted here was obtained from the 3D dust cells, and projected along the line of sight. The top row shows the physical maps of SFR, stellar mass ($M_\mathrm{star}$), and specific star-formation rate (sSFR = $\mathrm{SFR}/M_\mathrm{star}$). The stellar mass of both galaxies was estimated through the 3.6~$\mu$m luminosities, following the prescriptions of \citet{Oliver_2010MNRAS.405.2279O}. The morphological type of each galaxy was taken into account when calculating the $M_\mathrm{star}$. We convert the intrinsic FUV luminosity into SFR based on the recipe of \citet{Kennicutt_2012ARA&A..50..531K} for a \citet{Chabrier_2003PASP..115..763C} IMF. The bottom row of Fig.~\ref{fig:m51_physical_maps} presents the dust temperature ($T_\mathrm{dust}$), dust mass ($M_\mathrm{dust}$), dust density ($\rho_\mathrm{dust}$) maps. The diffuse dust temperature of each dust cell in our simulations was approximated through the strength of the ISRF $(U)$, following the method used in \citet{Nersesian_2020A&A...637A..25N}. Assuming that dust is heated by an ISRF with a Milky-Way like spectrum \citep{Mathis_1983A&A...128..212M} the dust temperatures of the diffuse dust can be approximated with

\begin{equation} \label{eq:Tdust}
T_\mathrm{dust} \ = T_\mathrm{o} \ U^{1/(4+\beta)} \, ,
\end{equation}   

\noindent \citep{Aniano_2012ApJ...756..138A, Nersesian_2019A&A...624A..80N}. $T_\mathrm{o}$ is defined as 18.3~K, the dust temperature measured in the solar neighbourhood, and $\beta = 1.79$ is the dust emissivity index, for the \textsc{THEMIS} dust model.

The top-left panel in Fig.~\ref{fig:m51_physical_maps} shows the distribution of SFR per pixel across M~51. Both the spiral arms and the central region of M~51a are more pronounced, highlighting the areas where new stars are brewing. Interestingly, we do not notice an extremely enhanced SFR in the upper body of M~51a. This result is in line with theory that suggests the recent interaction between the two galaxies, 50 to 100~Myr ago, not leading up to a major ongoing starburst. The ability of M~51a to produce stars was unaffected by that interaction with M~51b, and any signs of star formation are attributed to its normal activity. As a repercussion of the overestimated FUV luminosity in M~51b (Sect.~\ref{subsec:obs_vs_mod}), its nucleus is visible on this particular map, giving a false impression of the actual star-forming activity of the galaxy. Despite the overestimation of the SFR in M~51b, we are confident that it will not affect the results of this study as we only consider an old stellar population for M~51b.     

Comparing the SFR to the dust temperature map (or the strength of the ISRF), we notice some common features, i.e. the central regions of both galaxies as well as the spiral arms are more prominent here. We measure a weighted average of the global dust temperature for the entire system of $16.1\pm2.2$~K, whereas the corresponding dust temperature for M~51a is $16.1\pm1.6$~K, and for M~51b is $16.8\pm1.6$~K. In the central bulge region of both galaxies, we measure an average dust temperature of $20.2\pm1.2$~K and $19.6\pm1.3$~K, respectively. The dust temperature in the spiral arms of M~51a is of the same level as the dust temperature in the centre, pinpointing the sites of star formation. The inter-arm regions have a moderate dust temperature with typical values ranging from 13~K to 17~K, while the dust temperature takes its lowest values ($\sim10$~K) around the outer radii of the disc, where it is more unlikely for young stars to be formed. 

Consistent with the findings of \citet{Mentuch_Cooper_2012ApJ...755..165M}, the dust in the centre of M~51b experiences the effect of the dense ISRF, raising the dust temperature to relatively high levels. At first, these results might be thought of as contradictory, due to the apparent correlation between the dust temperature and the star-forming activity of a galaxy. However, several studies have shown that early-type galaxies (galaxies with their ISRF dominated by an evolved stellar population), such as M~51b, tend to have on average warmer dust temperatures than late-type galaxies \citep[e.g.][]{Skibba_2011ApJ...738...89S, Bendo_2012MNRAS.419.1833B, Boselli_2012A&A...540A..54B, Smith_2012ApJ...748..123S, Viaene_2017A&A...599A..64V, Nersesian_2019A&A...624A..80N}. \citet{Mentuch_Cooper_2012ApJ...755..165M} proposed an AGN (active galactic nucleus) as an alternative heating mechanism of the dust in M~51b. This assumption was based on the optical signature of AGN reported by \citet{Moustakas_2010ApJS..190..233M}. Nevertheless, recent cm and mm continuum observations in the nucleus of M~51b do not show evidence of a buried AGN activity \citep{Alatalo_2016ApJ...830..137A}. On the contrary, more and more evidence surfaced that support an evolved stellar population as the main dust heating source in M~51b \citep[e.g.][]{Alatalo_2016ApJ...830..137A, Eufrasio_2017ApJ...851...10E, Wei_2020arXiv200706231W}.

Another potential dust heating mechanism is by collisional heating with electrons in a hot gas. X-ray images from \textit{Chandra} revealed a weak X-ray emission in the nucleus of M~51b, and two X-ray arcs of cooling gas near the nucleus, most likely produced by feedback from the central supermassive black hole \citep{Schlegel_2016ApJ...823...75S}. \citet{Schlegel_2016ApJ...823...75S} measured the gas temperature of the nucleus to be 0.61~keV, and that of the two arcs $\sim0.4$ and 0.65~keV. The authors also showed the presence of an H$\alpha$ structure right beyond the X-ray arcs, suggesting that the expanding plasma pushed enough material outward, which could trigger the process of star formation. Finally, UV emission originating from M~51a may be an alternative heating source. But as we have shown in the previous section, the contribution provided by M~51a comes down to the percentage level. In reality, the combined effect of the discussed heating mechanisms could explain the increased dust temperature at the very centre of M~51b. Notwithstanding, our analysis shows that M~51b's native old stellar population is the main dust heating source.

Looking now at the stellar and dust mass maps, we see a smooth distribution for the former and a more structured distribution for the latter. For M~51a, there are some features highlighting the spiral structure in the $M_\mathrm{star}$ map, but they are not as pronounced as in the SFR, $T_\mathrm{dust}$ or $M_\mathrm{dust}$ maps. Similarly, the sSFR map shows even smoother distribution without any sign of structure in the disc, indicating a somewhat constant sSFR throughout the discs of M~51a ($2.36\times10^{-5}~\mathrm{Myr}^{-1}$), and M~51b ($2.70\times10^{-7}~\mathrm{Myr}^{-1}$). The sSFR can be considered as a measure of the current over the past SFR for a galaxy. The bulge region of M~51a has bluer colours indicating that next to no star formation is happening in the very centre of the galaxy, while the disc of M~51b is completely absent supporting the findings of other studies of a complete lack of recent SFR in M~51b. On the other hand, the dust density map highlights the fine structure of the spiral arms of M~51a. The dust density of M~51a peaks up around $3.26\pm0.02\times10^{-4}$~M$_\odot$~pc$^{-3}$ in the first 2~kpc radius, and then slightly decreases along the spiral arms. Around the edges and in the inter-arms regions we see a drop in dust density of about an order of magnitude, which further confirms our suspicion on why the photons from M~51b were able to propagate so deep in the disc of M~51a (see right panel of Fig.~\ref{fig:m51_heating_maps}). As expected, the sites of higher dust density match the regions of enhanced SFR, dust temperature and stellar mass. Finally, for M~51b we measure an average dust density of $8.45\pm0.07\times10^{-5}$~M$_\odot$~pc$^{-3}$ within a 2~kpc radius. This value is an order of magnitude lower than the one measured for the centre of M~51a, but it is comparable to the values measured for the dust density in the tidal bridge and outer spiral arms of M~51a. 

\begin{figure}[t!]
\centering
\includegraphics[width=\textwidth]{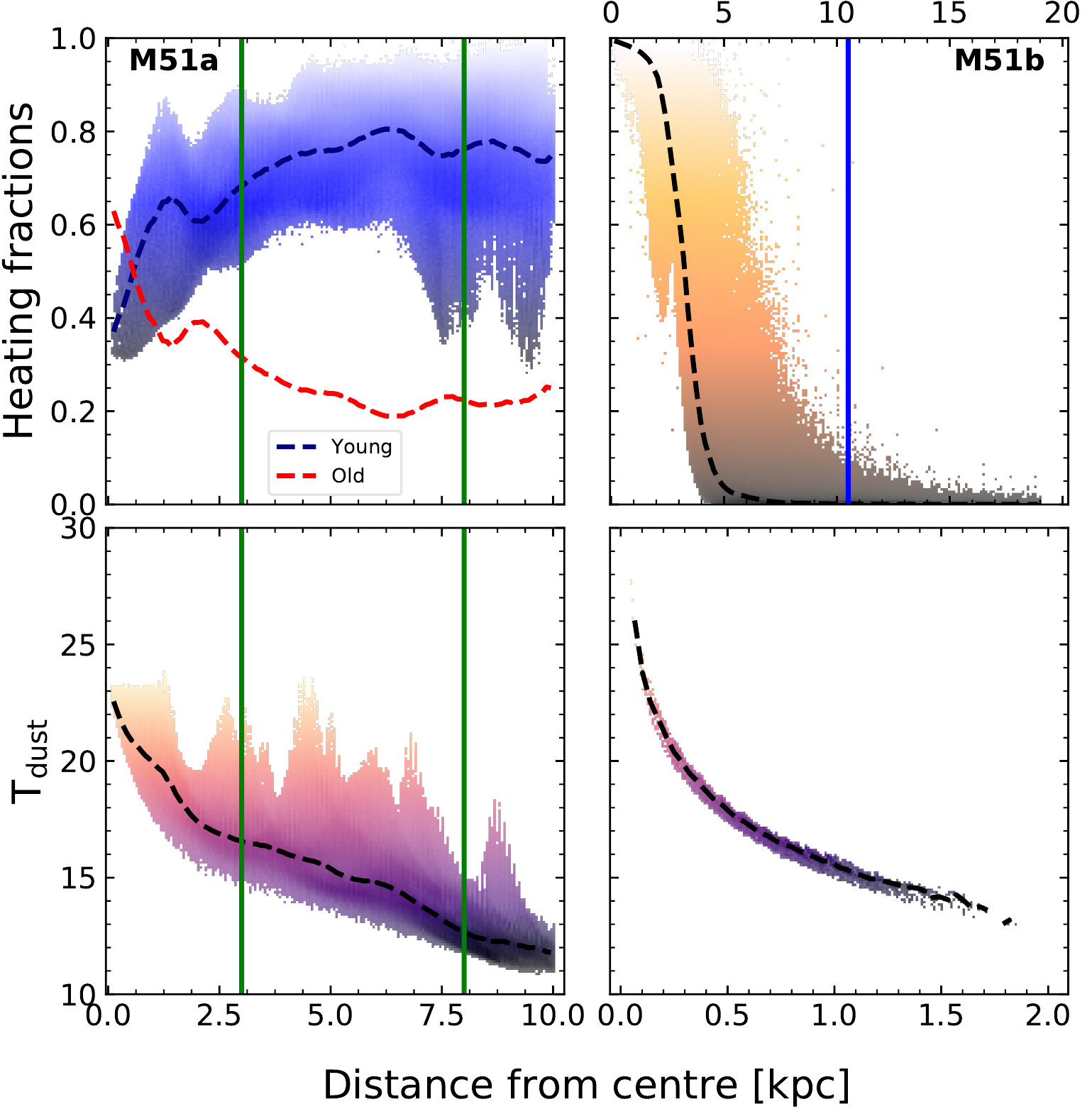}
\caption{Various radial profiles of M~51, as obtained from the 3D dust cell data. \textit{First column}: from top to bottom, the radial profiles of the dust heating fractions, and the dust temperature, for M~51a. The vertical green lines mark the 3~kpc and 8~kpc radii. \textit{Second column}: The top right panel shows the radial profile of $f_\text{M51b}$ up to 20~kpc radii from the centre of M~51b, and the blue line marks the location of the central region of M~51a. The bottom right panel shows the radial profile of the dust temperature of M~51b up to a radius of 2~kpc radii. The dashed lines passing though the points represent the running median. The points are colour coded according to the same quantity as in the \textit{y-axis}. The level of transparency indicates the points density.}
\label{fig:rad_profiles}
\end{figure}

\subsection{Radial profiles} \label{subsec:rad_prof}

Figure~\ref{fig:rad_profiles} shows the radial profiles of various physical quantities as a function of the galactocentric distance from M~51a (first column) and from M~51b (second column). The dashed lines passing through the data points are the running medians and show the general trend of the various radial profiles. The top row shows the heating fractions by the different stellar components in our model (see also Equation~\ref{eq:heating_fractions}). Specifically, the top panel of the first column shows the radial profiles of the heating fractions by the young (blue points and blue dashed line) and old (red dashed line) stellar populations in M~51a. Here we see that the central region (up to 1~kpc distance), is populated by a considerable fraction of old stars donating, on average, $43.2\%$ of their energy to the dust heating. Young stars dominate the radiation field of M~51a and are the predominant dust heating source in the galactic disc, with a global percentage of $72.1\%$. Following the running median, we see that the dust heating by the young (old) stellar population sharply rises (falls) within the 1~kpc, then drops (rises) to a local minimum (maximum), and then beyond the 3~kpc radius it stays roughly constant at $\sim76\%$ in the outer radii.

The top panel of the second column of Fig.~\ref{fig:rad_profiles} shows the extent of the radiative effect that M~51b has on the dusty disc of M~51a. The \textit{x-axis} is the galactocentric distance from M~51b up to 20~kpc radius. Within the first few kpcs, the contribution of the stellar radiation of M~51b to the energy distribution of M~51a is non-negligible. On the contrary, it is responsible for $38\%$ of the dust heating in the area between 2~kpc and 5~kpc from the centre of M~51b, which falls right where the upper spiral arms of M~51a are located. After the 5~kpc mark, the dust heating provided by M~51b drops to a mere $4.5\%$, while after a distance of 8~kpc remains below the percentage level. This is a surprising result which could have significant consequences on studies of the ISM properties in pre-mergers, where the primary galaxy is often in focus while the companion considered to have no relevant effect whatsoever. Here we prove that the companion galaxy, even in an early merging phase, as M~51, could significantly alter the energy budget of the main galaxy, and therefore cause erroneous results if not accounted for. 

The last row of Fig.~\ref{fig:rad_profiles} showcases the dust temperature of the diffuse dust in each galaxy versus the galactocentric distance. For M~51a, we observe a high dust temperature in the centre up to 23.8~K which slowly drops toward the outer regions. Several peaks of higher dust temperatures can be seen along the radii which indicate the location of the star-forming regions across the spiral arms (see the $T_\mathrm{dust}$ map in Fig.~\ref{fig:m51_physical_maps}). The outer regions and edges of the disc are occupied by the coldest diffuse dust in the galaxy, with average temperatures of $12.4\pm0.8$~K. On the other hand, M~51b shows a smooth dust distribution with the highest temperature in the very centre being 27.7~K, $\sim4$~K higher than the dust temperature measured in the centre of M~51a. Similarly, \citet{Mentuch_Cooper_2012ApJ...755..165M} measured a peak in dust temperature of 30~K from single modified blackbody fits to the 70 -- 500~$\mu$m images of M~51b. As a result of the extremely dense ISRF in the bulge region of M~51b, dust can be warmed up to such extremely high temperatures. Similarly, when investigating the dust heating in the bulge of the Andromeda galaxy it was found that the old stellar population is responsible for the highest dust temperatures in the galaxy \citep{Groves_2012MNRAS.426..892G, Viaene_2014A&A...567A..71V}. Beyond the nucleus, the dust temperature of M~51b rapidly drops down to $16.6\pm0.8$~K in the first kpc, while for the next kpc remains roughly constant with an average dust temperature of $15.0\pm0.6$~K. The trend in the dust temperature, is another indication that the diffuse dust density decreases as we move further away from the highly dense nucleus.

\section{Discussion} \label{sec:discussion}

The investigation of UV-to-optical colour-magnitude diagrams (e.g. NUV$-r$ vs $M_r$) of galaxies revealed that they can be separated into two distinct populations: the `red sequence', that is occupied by passive galaxies with red colours, and the `blue cloud' populated by actively star-forming galaxies of blue colours \citep[e.g.][and references therein]{Wyder_2007ApJS..173..293W, Salim_2007ApJS..173..267S}. This bimodality in the UV-optical colours does appear to be linked with morphology as well. Blue galaxies tend to have a prominent disc structure, while red galaxies tend to be more massive and to exhibit a smooth ellipsoidal shape without signs of a disc or spiral arms \citep{Gadotti_2009MNRAS.393.1531G, Wuyts_2011ApJ...742...96W, Whitaker_2015ApJ...811L..12W}. Further analysis on the global properties of galaxies confirmed the existence of a `main sequence of star-forming galaxies' (MS) in the SFR-$M_\mathrm{star}$ plane, as well as a region occupied by passive galaxies \citep{Noeske_2007ApJ...660L..43N, Elbaz_2007A&A...468...33E, Daddi_2007ApJ...670..156D}. Inevitably, someone would wonder if such a dichotomy arises from the internal processes that govern the global properties of galaxies. Spatial studies in H\textsc{ii} regions of nearby galaxies \citep[e.g.][]{Rosales_Ortega_2012ApJ...756L..31R, Sanchez_2013A&A...554A..58S}, as well as in multi-wavelength broadband imaging \citep[e.g.][]{Wuyts_2013ApJ...779..135W, Enia_2020MNRAS.493.4107E, Morselli_2020MNRAS.tmp.1946M}, showed that the relation between the stellar mass and SFR surface densities holds up on kpc and sub-kpc scales. In this section, we analyse the spatially resolved distributions of $\mu_\star$ and $\Sigma_\text{SFR}$ as encrypted within the pixels of the corresponding maps in Fig.~\ref{fig:m51_physical_maps}. M~51 is the ideal system to explore the $\Sigma_\text{SFR}$--$\mu_\star$ plane. That is because M~51a is a late-type galaxy of moderate SFR, whereas M~51b resembles more of a `red and dead' galaxy, where the SFR activity has ceased altogether. The correlation between $f_\text{young}$ and sSFR is discussed as well.    

\subsection{$\Sigma_\text{SFR}$--$\mu_\star$ relation for M~51}

\begin{figure}[t!]
\centering
\includegraphics[width=\textwidth]{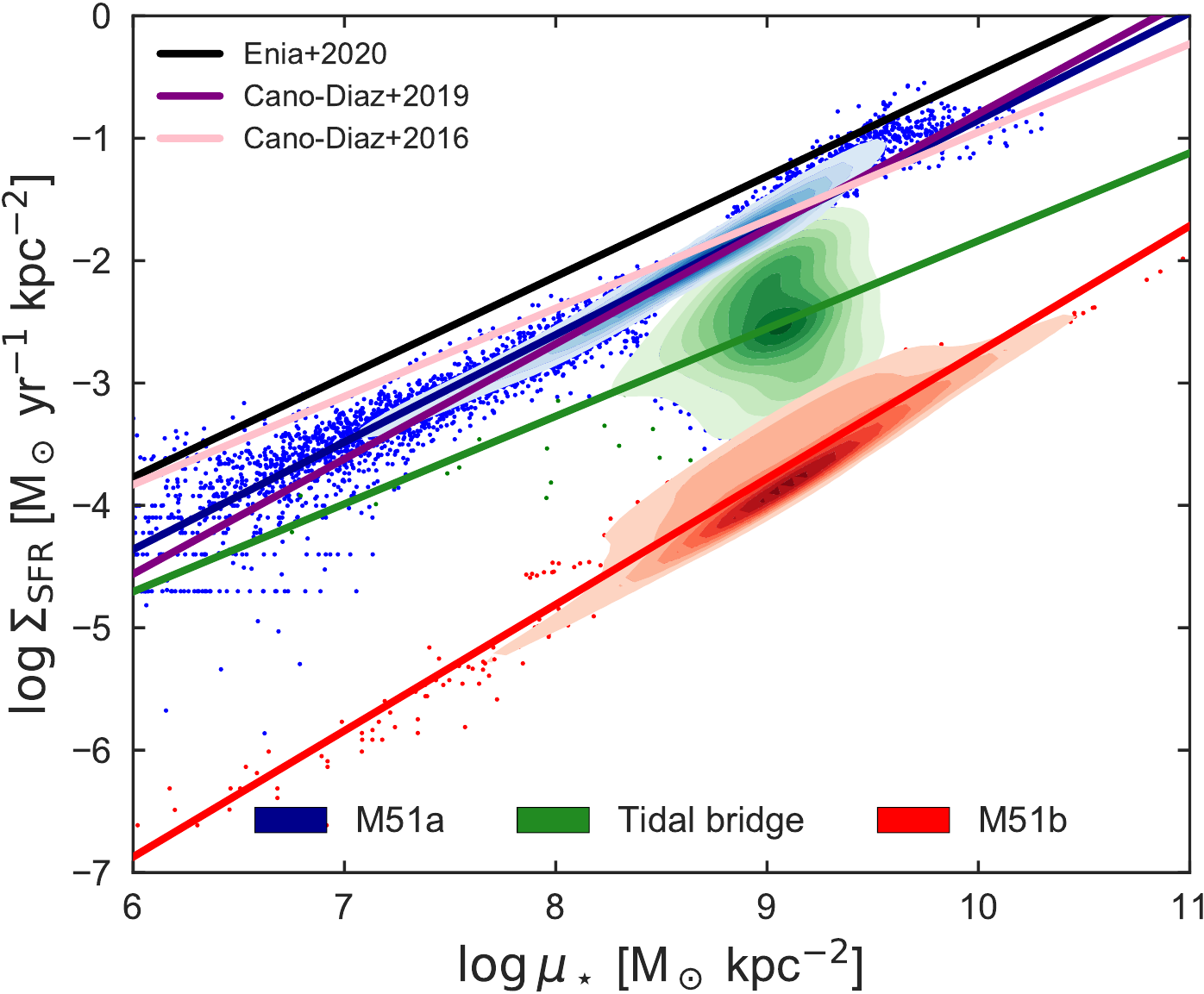}
\caption{Plot of $\Sigma_\text{SFR}$ against stellar mass density for M~51 (physical scale of 153.4~pc). The blue density distribution represents the pixels that belong to M~51a, the green distribution corresponds to the pixels of the tidal bridge, and the red distribution denotes the pixels of M~51b within a 2~kpc radius. The black line is the relation presented in \citep{Enia_2020MNRAS.493.4107E}, the purple line is the best fit relation from MaNGA \citep{Cano_Diaz_2019MNRAS.488.3929C}, and the pink line for galaxies in the CALIFA survey \citep{Cano_Diaz_2016ApJ...821L..26C}. The remaining coloured lines are a fit through the distribution of the same colour.}
\label{fig:m51_ms}
\end{figure}

In Fig.~\ref{fig:m51_ms}, we plot $\Sigma_\text{SFR}$ against the stellar mass density $\mu_\star$, for a physical scale of 153.4~pc. The blue and red KDE distributions belong to all pixels related to M~51a, and to M~51b within a 2~kpc radius from its centre, respectively. The green distribution depicts the relation for the tidal bridge\footnote{The region of the tidal bridge is defined as in \citet{Lee_2012ApJ...754...80L}.} that connects both galaxies. A linear fit for each distribution was performed yielding the following relations,

\begin{equation*}
\resizebox{0.93\hsize}{!}{%
$\log\left(\Sigma_\text{SFR}/\text{M}_\odot~\text{yr}^{-1}~\text{kpc}^{-2}\right) = 0.88 \log\left(\mu_\star/\text{M}_\odot~\text{kpc}^{-2}\right) -9.64\, \left[\text{M~51a}\right]$%
}
\end{equation*}
\begin{equation}\label{eq:best_fitting_values}
\resizebox{0.93\hsize}{!}{%
$\log\left(\Sigma_\text{SFR}/\text{M}_\odot~\text{yr}^{-1}~\text{kpc}^{-2}\right) = 0.72 \log\left(\mu_\star/\text{M}_\odot~\text{kpc}^{-2}\right) -9.01\, \left[\text{TB}\right]$%
}
\end{equation}
\begin{equation*}
\resizebox{0.93\hsize}{!}{%
$\log\left(\Sigma_\text{SFR}/\text{M}_\odot~\text{yr}^{-1}~\text{kpc}^{-2}\right) = 1.03 \log\left(\mu_\star/\text{M}_\odot~\text{kpc}^{-2}\right) -12.88\, \left[\text{M~51b}\right].$%
}
\end{equation*}

\noindent As a reference, we over-plot the spatially resolved `main sequence' derived by \citet[][black line]{Enia_2020MNRAS.493.4107E} for eight DustPedia spiral galaxies (including M~51a), on spatial scales of $8\arcsec$ or 0.2-0.8~kpc depending on the distance of each galaxy. They obtained $\log\left(\Sigma_\text{SFR}\right)=0.82\log\left(\mu_\star\right)-8.69$. Furthermore, we show the relation fitted by \citet[][purple line]{Cano_Diaz_2019MNRAS.488.3929C} to $\sim2000$ galaxies from the MaNGA MPL-5 survey \citep{Bundy_2015ApJ...798....7B}, on spatial scales of $\sim1.0$~kpc and in the redshift range $0.01 < z < 0.15$. This line has a slope of 0.94. Finally, \citet{Cano_Diaz_2016ApJ...821L..26C} retrieved a slope of 0.72 (pink line) for a sample of 306 galaxies from the CALIFA survey \citep{Sanchez_2012A&A...538A...8S}, at $0.005 < z < 0.03$, and on spatial scales of 0.5-1.5 kpc. All relations have been converted to a \citet{Chabrier_2003PASP..115..763C} IMF.

Several studies of nearby galaxies have worked at a kpc scale or slightly lower, in order to address if the $\Sigma_\text{SFR}$ to gas surface density relation holds on sub-galactic scales \citep[e.g.][]{Bigiel_2008AJ....136.2846B, Leroy_2008AJ....136.2782L, Leroy_2012AJ....144....3L, Casasola_2015A&A...577A.135C}. In an effort to study how small-scale ISM structure relates to the ability of gas to form stars in M~51a, \citet{Leroy_2017ApJ...846...71L} used a spatial scale of 370~pc to avoid evolutionary effects of individual regions, and found fluctuations between 370~pc and 1.1~kpc scales. \citet{Onodera_2010ApJ...722L.127O} found a large dispersion in the $\Sigma_\text{SFR}$ inside M~33 at a resolution of 80~pc, that they attribute to various evolutionary stages of the giant molecular clouds. Recipes based on energy budget have even been more challenging at sub-kpc scales. \citet{Williams_2019MNRAS.487.2753W} showed that the local dust-energy balance does not hold below scales of 1.5~kpc, while \citet{Boquien_2015A&A...578A...8B} showed that below 1.0~kpc hybrid SFR estimators (also based on an energy budget) are not all consistent. In our model we use a spatial scale of $\sim153$~pc similar to the characteristic size of a large H\textsc{ii} region \citep[e.g.][]{Rousseau_Nepton_2018MNRAS.477.4152R}, which may challenge the definition of an SFR calibration. Consequently, we tested the consistency of the MS of M~51 on different spatial scales. Instead of running new simulations with lower resolution, we regrid the output to two additional scales of 460~pc and 1500~pc, and then apply the SFR and $M_\text{star}$ recipes to the data. The fitted slopes of M~51a and M~51b remain roughly the same across the different spatial scales. For M~51a we retrieve a slope of 0.92 and 0.89, while for M~51b we obtain a slope of 0.92 and 1.07, for the respective scales. Since all three representations of M~51's main sequence have similar slopes despite the change in the physical scales, makes us confident that the used recipes can be applied on scales at least up to 150~pc. 

The total dispersion of M~51a’s MS is $\sigma=0.10$, for M~51b is $\sigma=0.23$, and for the tidal bridge $\sigma=0.15$. The scatter of M~51a is marginally consistent with the typical range of 0.13-0.35 reported in previous studies on the spatially resolved MS \citep{Magdis_2016MNRAS.456.4533M, Maragkoudakis_2017MNRAS.466.1192M, Hsieh_2017ApJ...851L..24H, Hall_2018ApJ...865..154H, Enia_2020MNRAS.493.4107E}. We also investigate if the scatter of the MS varies with mass by splitting the data in three mass bins: $6 \le \log \mu_\star < 7.5$, $7.5 \le \log \mu_\star < 9$, and $\log \mu_\star \ge 9$. For M~51a we find a low dispersion in the first two bins of $\sigma \sim 0.07$ and an increased scatter for the highest mass bin of $\sigma=0.13$. For M~51b we find an increasing dispersion with $\log \mu_\star$, that is $\sigma=0.06$, $\sigma=0.21$, and $\sigma=0.24$, respectively for each bin. We do not provide the dispersion per bin for the tidal bridge due to the reduced number of points in the lowest and highest mass bins.

Interestingly enough, the pixel distribution of each galaxy occupies two distinct regions in the $\Sigma_\text{SFR}$--$\mu_\star$ plane. Their position in Fig.~\ref{fig:m51_ms} is purely a consequence of their SFH. The distribution of M~51a falls predominately on what is described as the `main sequence' of star-forming galaxies. Indeed, the slope we retrieve for M~51a is similar to the MS obtained by \citet{Enia_2020MNRAS.493.4107E}, but a vertical shift is evident possibly due to the differet methods of estimating the SFR and $M_\text{star}$. Although, the slopes for the CALIFA and MaNGA surveys differ from the one of M~51a, all share, more or less, the same locus. The steep slope of the observed `main sequence' indicates the change in $\Sigma_\text{SFR}$ per $\mu_\star$ in the different regions of M~51a, ranging approximately by three orders of magnitude. On the other hand, the distribution related to M~51b reside on a well defined `sequence' but with lower (2--3 orders of magnitude) values of $\Sigma_\text{SFR}$ than that of M~51a. The `red sequence' has a steeper slope, with $\Sigma_\text{SFR}$ varying about five orders of magnitude. The tidal bridge seems to have a well defined star-forming locus in the plane of Fig.~\ref{fig:m51_ms}, possibly as a result of the tidal interaction. The stellar bridge not only connects the two galaxies in physical space, but also along the $\Sigma_\text{SFR}$--$\mu_\star$ relation. For a fixed $\mu_\star$, the $\Sigma_\text{SFR}$ of the tidal bridge is always lower than the $\Sigma_\text{SFR}$ of M~51a, and larger than the $\Sigma_\text{SFR}$ of M~51b. This indicates that the SFR in that region was once enhanced during the passage of M~51b, through the exchange of gas between the two components, enriching one of them and depleting the other, and it is now going towards a quenching phase. However, as stated in \citet{Lee_2012ApJ...754...80L} it is still unclear whether the young stars populating the bridge were born in situ or migrated from the spiral arm of M~51a. 

\begin{figure}[t!]
\centering
\includegraphics[width=\textwidth]{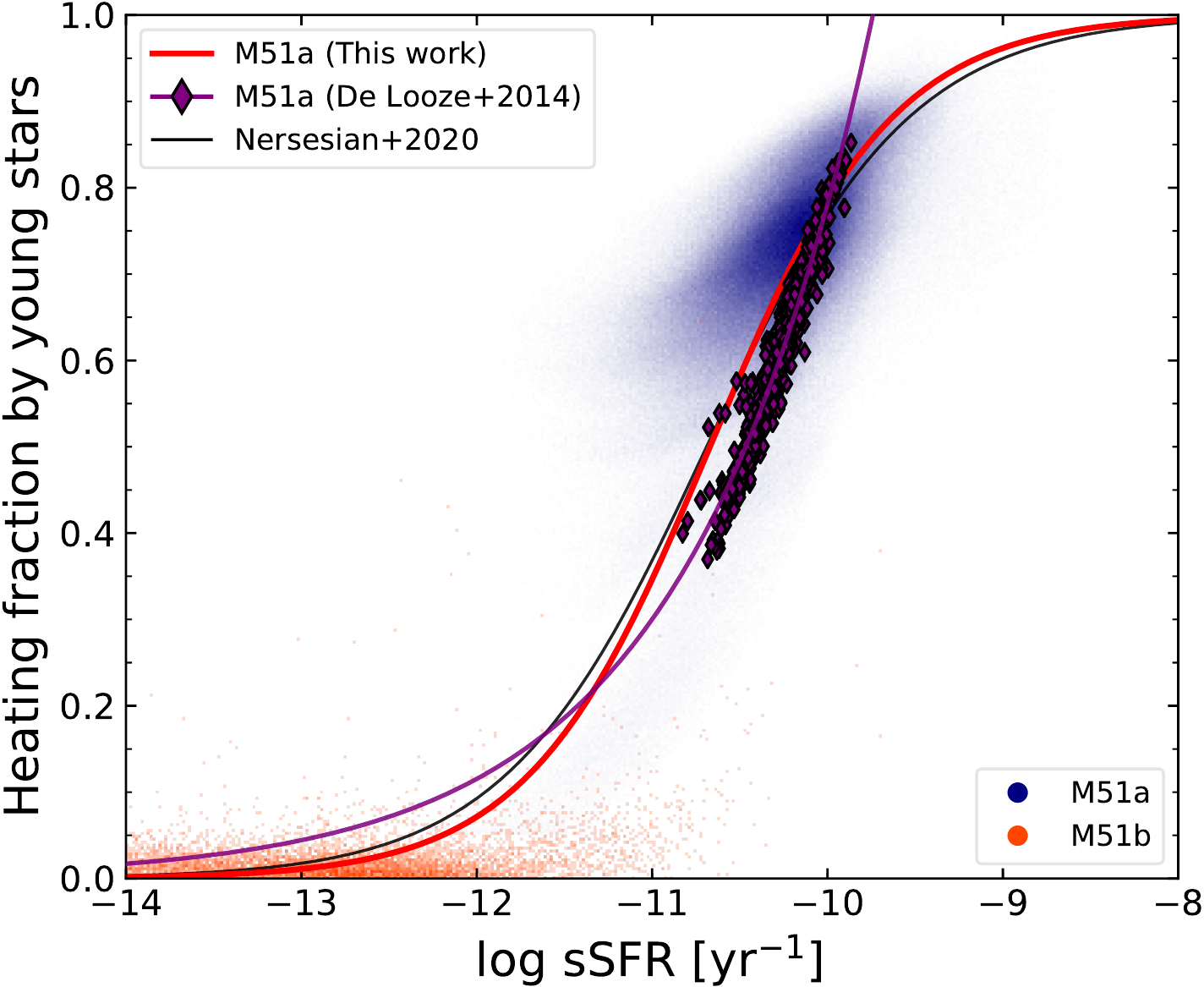}
\caption{Relation between sSFR and $f_\text{young}$. The blue points represent the dust cells that belong to M~51a within a 10~kpc radius from its centre, and the orange points the dust cells of M~51b. The level of transparency indicates the points density. The red line is the fit through the points of M~51a with Equation~\ref{eq:tanh}. The purple diamonds are the pixel values of M~51a, and the purple line is the power-law fit to that data from the model of \citet{De_Looze_2014A&A...571A..69D}. The black line shows the `main sequence' of galaxies as derived in \citet{Nersesian_2020A&A...637A..25N} by fitting the voxels of the radiative transfer models of six face-on late-type galaxies. }
\label{fig:m51_ssfr}
\end{figure}

\subsection{Correlation between young heating fraction and sSFR}

Another tight correlation that reveals the spatial variation of star formation is the one between the young heating fraction and the sSFR \citep{De_Looze_2014A&A...571A..69D, Viaene_2016A&A...586A..13V, Viaene_2017A&A...599A..64V, Verstocken_2020A&A...637A..24V, Nersesian_2019A&A...624A..80N, Nersesian_2020A&A...637A..25N}. In \citet{Nersesian_2020A&A...637A..25N}, the authors extended the parameter range of the relation \citep[first presented by][for M~51a]{De_Looze_2014A&A...571A..69D}, by concentrating the results from the radiation transfer models of six late-type, face-on spiral galaxies, and derived a sequence for sub-kpc regions within those galaxies. Figure~\ref{fig:m51_ssfr} depicts the young heating fraction against the sSFR of M~51a (blue points), and extends the parameter space of the relation by including the data of a non-star-forming galaxy, M~51b (orange points). We also over-plot the pixel data values (purple diamonds), as well as the best fitted power-law (purple solid line) from the radiative transfer model of M~51a by \citet{De_Looze_2014A&A...571A..69D}. Finally, Fig.~\ref{fig:m51_ssfr} shows the sequence as derived by \citet{Nersesian_2020A&A...637A..25N}, using the following equation

\begin{equation} \label{eq:tanh}
f_\text{young} = \frac{1}{2} \left[1 - \tanh\left(a \log\left(\text{sSFR}/\text{yr}^{-1}\right) + b\right)\right] \, .
\end{equation}

\noindent The same equation is used to fit the relation for M~51a (red line). The fitting yields $a = -0.96$ and $b = -10.30$. Comparing the data from the two different radiative transfer models of M~51a, it is evident that they are in very good agreement, residing within the same parameter space, but with a slight offset towards lower sSFR for our model. This small offset is attributed to the different SFR recipes used by the two studies. Overall, the main body of data points of M~51a reside in the middle region of the solid black line, while the orange cloud of points related to M~51b is located at the lower-left end of the same relation. In contrast to Fig.~\ref{fig:m51_ms}, where the data from both galaxies were separated into two distinct regions, here we see that there is a continuous change in the properties (and in morphology) of the two galaxies along the sequence. 

\section{Summary \& Conclusions} \label{sec:conclusions}

We have constructed a detailed 3D radiative transfer model for the interacting system M~51, using the state-of-the-art Monte Carlo code \textsc{SKIRT}. The scope of this study is to investigate the dust-heating processes in a merging system, and to assess the radiative influence of the companion, on the energy budget of its parent galaxy. As a continuation of this series of papers, started with the study of M~51a \citep{De_Looze_2014A&A...571A..69D}, we also present an updated model of the grand-design spiral galaxy M~51a based on the methodology presented in \citet{Verstocken_2020A&A...637A..24V}. Our model has been validated by comparing the simulated SEDs with the observational data across the entire UV to submm wavelength range, yielding a best-fitting description of the interacting system. Here we list our main findings:

\begin{itemize}

\item 40.7\% of the bolometric luminosity is absorbed by dust in M~51. This fraction is in line with the mean values determined by \citet{Bianchi_2018A&A...620A.112B}, for late-type spiral galaxies of morphological type Sb-Sc.

\item While the young stellar population is the main dust heating agent of M~51, the older stellar population is also having a quite active role in the process of dust heating. The global $f_\text{young}$ ($f_\text{old}$) fraction of M~51 is $71.2\%$ ($23\%$), while the remaining $5.8\%$ is provided by M~51b. By isolating each galaxy, we measure that M~51a's diffuse dust is heated by $f_\text{young}$ ($f_\text{old}$) of $72.1\%$ ($23.1\%$) and a $f_\text{M51b}$ of $4.8\%$. The respective heating fractions for M~51b's dust are $f_\text{young}$+$f_\text{old} = 2\%$, and a $f_\text{M51b}$ of $98\%$.

\item We find that the extent of the contribution of the dust heating by the companion galaxy depends on the density of the ISM, with photons easily propagating to the less dense outskirts of M~51a’s disc through scattering, while they are absorbed by the denser regions in the upper body of M~51a. Specifically the contribution of the radiation field of M~51b is significant up to a distance of 5~kpc from its centre, with a heating fraction of $\sim38\%$. After the 5~kpc mark the significance of this heating agent drops below 10\% and remains under the percentage level beyond a radius of 8~kpc. 

\item On average, the diffuse dust temperature at the central region of both galaxies is warmer than the dust temperatures found at their discs, with $T_\mathrm{dust}$ decreasing with galactocentric distance. The young stellar population is responsible for the warmer $T_\mathrm{dust}$ in the spiral arms and near the star-forming dust clouds of M~51a, whereas the dense old ISRF in the nucleus of M~51b, raises the dust temperature to relatively high levels. The weighted average dust temperature of M~51a is $16.1\pm1.6~$K, and for M~51b $16.8\pm1.8~$K.

\item We further confirm that the pixel distribution related to each galaxy falls into two distinct regions in the $\Sigma_\text{SFR}$--$\mu_\star$ plane (physical scale of 153.4~pc). The data from M~51a are located in the region of the diagram that resembles the `main sequence of star-forming galaxies', while the data from M~51b occupy the parameter space known as the `red sequence' of non-star-forming galaxies.

\item Exploring the $f_\text{young}$--sSFR plane \citep{De_Looze_2014A&A...571A..69D, Nersesian_2020A&A...637A..25N}, we find that instead of two distinct galaxy populations, galaxies follow a continuous relation depending on their intrinsic properties and morphology.

\end{itemize}

Our results, model configuration ({\tt{ski}} file), and model input data (component maps) are publicly available on the DustPedia archive (\url{http://dustpedia.astro.noa.gr}) under the \textsc{SKIRT} tab, so that anyone can reproduce our work. 

\begin{acknowledgements}
We would like to thank the anonymous referee for providing detailed comments and suggestions, which helped to improve the presentation of the manuscript. AN gratefully acknowledges the support by Greece and the European Union (European Social Fund-ESF) through the Operational Programme "Human Resources Development, Education and Lifelong Learning 2014-2020" in the context of the project “Anatomy of galaxies: their stellar and dust content through cosmic time” (MIS 5052455). DustPedia is a collaborative focused research project supported by the European Union under the Seventh Framework Programme (2007-2013) call (proposal no. 606847). The participating institutions are: Cardiff University, UK; National Observatory of Athens, Greece; Ghent University, Belgium; Universit\'{e} Paris Sud, France; National Institute for Astrophysics, Italy and CEA, France. The computational resources (Stevin Supercomputer Infrastructure) and services used in this work were provided by the VSC (Flemish Supercomputer Center), funded by Ghent University, FWO and the Flemish Government – department EWI. This research made use of Astropy,\footnote{\url{http://www.astropy.org}} a community-developed core Python package for Astronomy \citep{Astropy_2013A&A...558A..33A, Astropy_2018AJ....156..123A}. This research has made use of the NASA/IPAC Infrared Science Archive (IRSA; \url{http://irsa.ipac.caltech.edu}), and the NASA/IPAC Extragalactic Database (NED; \url{https://ned.ipac.caltech.edu}), both of which are operated by the Jet Propulsion Laboratory, California Institute of Technology, under contract with the National Aeronautics and Space Administration. 
\end{acknowledgements}

\bibliographystyle{aa}
\bibliography{References}
%
%
\appendix
%

\section{Global photometry}\label{ap:photometry}

Table~\ref{tab:phot} summarises the final flux densities and their corresponding uncertainties extracted from the global aperture photometry of the different galaxy regions listed in Table~\ref{tab:regions} and shown in Fig.~\ref{fig:photometry_M51}. The global flux densities were fitted with the SED code \textsc{CIGALE} (see Sect.~\ref{sec:cigale}), and were used to constrain the radiative transfer model of M~51 (see Sect.~\ref{sec:results}). Marked in boldface are the bands that were not used in our modelling.


\begin{table}[!ht]
\caption{Integrated flux densities for our galaxy sample in this paper, listed by increasing central wavelength. The bands not used in our modelling are indicated in boldface.}
\begin{center}
\scalebox{0.6}{
\begin{tabular}{lcccccc}
\hline 
\hline 
& & & & NGC~5194 (M~51a) & NGC5195 (M~51b) & M~51 \\
Instrument & Band & $\lambda_\mathrm{eff}$ & Pixel scale & Flux density & Flux density & Flux density \\
& & [$\mu$m] & [arcsec] & [Jy] & [Jy] & [Jy] \\
\hline
GALEX	&	FUV	&	0.154	& 3.2  &	$0.118	\pm	0.006$	&	$0.002	\pm	0.001$	&	$0.119	\pm	0.006$	\\
GALEX	&	NUV	&	0.227	& 3.2  &	$0.194	\pm	0.005$	&	$0.005	\pm	0.001$	&	$0.198	\pm	0.006$	\\
SDSS	&	$u$	&	0.359	& 0.45 &	$0.488	\pm	0.006$	&	$0.074	\pm	0.001$	&	$0.537	\pm	0.007$	\\
SDSS	&	$g$	&	0.464	& 0.45 &	$1.354	\pm	0.011$	&	$0.278	\pm	0.002$	&	$1.603	\pm	0.013$	\\
SDSS	&	$r$	&	0.612	& 0.45 &	$2.160	\pm	0.017$	&	$0.584	\pm	0.005$	&	$2.702	\pm	0.022$	\\
KPNO    &H$\alpha$& 0.657   & 0.3  &    $2.943  \pm 0.230$  &   $0.470  \pm 0.148$  &   $3.450  \pm 0.023$  \\
SDSS	&	$i$	&	0.744	& 0.45 &	$2.571	\pm	0.018$	&	$0.876	\pm	0.006$	&	$3.641	\pm	0.025$	\\
SDSS	&	$z$	&	0.89	& 0.45 &	$3.060	\pm	0.024$	&	$1.110	\pm	0.009$	&	$4.197	\pm	0.034$	\\
2MASS	&	$J$	&	1.235	& 1.0  &	$3.538	\pm	0.099$	&	$1.659	\pm	0.046$	&	$5.571	\pm	0.156$	\\
2MASS	&	$H$	&	1.662	& 1.0  &	$3.828	\pm	0.107$	&	$2.050	\pm	0.057$	&	$5.754	\pm	0.161$	\\
2MASS	&	$K_s$&	2.159	& 1.0  &	$3.446	\pm	0.097$	&	$1.660	\pm	0.046$	&	$5.295	\pm	0.148$	\\
WISE	&	W1	&	3.352	& 1.37 &	$2.368	\pm	0.069$	&	$0.885	\pm	0.026$	&	$3.260	\pm	0.095$	\\
IRAC	&	I1	&	3.508	& 0.75 &	$2.515	\pm	0.075$	&	$0.879	\pm	0.026$	&	$3.405	\pm	0.102$	\\
IRAC	&	I2	&	4.437	& 0.75 &	$1.608	\pm	0.048$	&	$0.561	\pm	0.017$	&	$2.286	\pm	0.069$	\\
WISE	&	W2	&	4.603	& 1.37 &	$1.408	\pm	0.048$	&	$0.516	\pm	0.018$	&	$2.076	\pm	0.071$	\\
IRAC	&	I3	&	5.628	& 0.6  &	$5.146	\pm	0.154$	&	$0.666	\pm	0.020$	&	$5.894	\pm	0.177$	\\
IRAC	&	I4	&	7.589	& 0.6  &	$12.899	\pm	0.387$	&	$1.037	\pm	0.031$	&	$14.476	\pm	0.434$	\\
WISE	&	W3	&	11.56	& 1.37 &	$8.520	\pm	0.392$	&	$0.851	\pm	0.039$	&	$9.837	\pm	0.452$	\\
WISE	&	W4	&	22.09	& 1.37 &	$13.167	\pm	0.737$	&	$1.578	\pm	0.088$	&	$15.164	\pm	0.849$	\\
MIPS	&	24	&	23.21	& 1.5  &	$11.859	\pm	0.593$	&	$1.537	\pm	0.077$	&	$13.546	\pm	0.677$	\\
\textbf{MIPS}	&	\textbf{70}	&	68.44	& 4.5  &	$128.945\pm	12.895$	&	$16.338	\pm	1.634$	&	$144.489\pm	14.449$	\\
PACS	&	70	&	68.92	& 2.0  &	$174.587\pm	12.221$	&	$22.356	\pm	1.565$	&	$205.312\pm	14.372$	\\
\textbf{MIPS}	&	\textbf{160}&	152.6	& 9.0  &	$391.906\pm	47.029$	&	$28.822	\pm	3.459$	&	$431.668\pm	51.800$	\\
PACS	&	160	&	153.9	& 4.0  &	$403.318\pm	28.232$	&	$28.642	\pm	2.005$	&	$429.377\pm	30.056$	\\
SPIRE	&	PSW	&	247.1	& 6.0  &	$189.055\pm	10.398$	&	$12.765	\pm	0.702$	&	$197.990\pm	10.889$	\\
SPIRE	&	PMW	&	346.7	& 8.0  &	$77.370	\pm	4.255$	&	$5.109	\pm	0.281$	&	$79.276	\pm	4.360$	\\
SPIRE	&	PLW	&	496.1	& 12.0 &	$26.061	\pm	1.433$	&	$1.697	\pm	0.093$	&	$25.907	\pm	1.425$	\\
\hline \hline
\end{tabular}}
\\[10pt]
\label{tab:phot}
\end{center}
\end{table}

Table~\ref{tab:comp_maps} gives an overview of the images that are used to produce the spatial distribution of the various components in our radiative transfer model of M~51, as well as the simulation output maps that are used to derive the local (per pixel) physical parameters.

\begin{table}[!ht]
\caption{Summary of the images used to produce the component maps in our radiative transfer model of M~51, as well as the simulation output images that are used to derive the local physical parameters. Indicated in bold are the image with the lowest resolution, determining the resolution of each map.}
\begin{center}
\scalebox{0.6}{
\begin{tabular}{lr}
\hline 
\hline 
Component map  & Observed images used\\
\hline
Old stellar disc        & \textbf{IRAC~3.6~$\mu$m}\\
Young non-ionising disc & GALEX~FUV, IRAC~3.6~$\mu$m, SDSS~$r$, MIPS~24~$\mu$m, PACS~70~$\mu$m, \textbf{PACS~160~$\mu$m}\\
Young ionising disc     & \textbf{MIPS~24~$\mu$m}, IRAC~3.6~$\mu$m, H$\alpha$\\
Dust disc               & GALEX~FUV, SDDS~$r$, MIPS~24~$\mu$m, PACS~70~$\mu$m, \textbf{PACS~160~$\mu$m}\\
\hline 
\hline 
Physical map  & Model images used\\
\hline
SFR             & GALEX~FUV, MIPS~24~$\mu$m, PACS~70~$\mu$m, \textbf{PACS~160~$\mu$m}\\
$M_\text{star}$ & IRAC~3.6~$\mu$m\\
\hline \hline
\end{tabular}}
\\[10pt]
\label{tab:comp_maps}
\end{center}
\end{table}

\end{document}